\documentclass[journal=jctcce,manuscript=article,layout=onecolumn]{achemso}

\usepackage[version=3]{mhchem} 
\usepackage{chemformula}
\usepackage{subcaption}
\usepackage{array}
\usepackage{enumitem}
\usepackage[frozencache,cachedir=minted-cache]{minted}
\usepackage{fancyvrb}
\usepackage{amsmath}
\usepackage{amssymb}

\usepackage[hidelinks]{hyperref}

\author{Joe Gilkes}
\affiliation[Chem]{Department of Chemistry, University of Warwick, Gibbet Hill Road, CV4 7AL Coventry, UK}
\alsoaffiliation[HetSys]{EPSRC HetSys Centre for Doctoral Training, University of Warwick, Gibbet Hill Rd, CV4 7AL, Coventry, UK}
\author{Mark Storr}
\affiliation[AWE]{AWE Plc, Aldermaston, UK}
\author{Reinhard J. Maurer}
\affiliation[Chem]{Department of Chemistry, University of Warwick, Gibbet Hill Road, CV4 7AL Coventry, UK}
\alsoaffiliation[Phys]{Department of Physics, University of Warwick, Gibbet Hill Road, CV4 7AL Coventry, UK}
\email{r.maurer@warwick.ac.uk}
\author{Scott Habershon}
\email{s.habershon@warwick.ac.uk}
\affiliation[Chem]{Department of Chemistry, University of Warwick, Gibbet Hill Road, CV4 7AL Coventry, UK}

\title[Kinetica.jl]
{
     Predicting long timescale kinetics under variable experimental conditions with Kinetica.jl  
}

\abbreviations{CRN, RRE, TS}

\begin{document}

\begin{center}
    UK Ministry of Defence \copyright \ Crown Owned Copyright 2024/AWE
\end{center}

\begin{tocentry}

\includegraphics{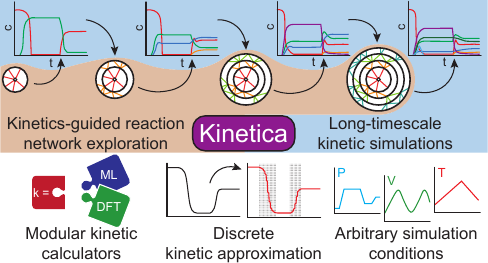}

\end{tocentry}

\newpage
\begin{abstract}
\singlespace
  \noindent Predicting the degradation processes of molecules over long timescales is a key aspect of industrial materials design. However, it is made computationally challenging by the need to construct large networks of chemical reactions that are relevant to the experimental conditions that kinetic models must mirror, with every reaction requiring accurate kinetic data. Here we showcase \textit{Kinetica.jl}, a new software package for constructing large-scale chemical reaction networks in a fully-automated fashion by exploring chemical reaction space with a kinetics-driven algorithm; coupled to efficient machine-learning models of activation energies for sampled elementary reactions, we show how this approach readily enables generation and kinetic characterization of networks containing $\sim10^{3}$ chemical species and $\simeq10^{4}$ -- $10^{5}$ reactions. Symbolic-numeric modelling of the generated reaction networks is used to allow for flexible, efficient computation of kinetic profiles under experimentally-realizable conditions such as continuously-variable temperature regimes, enabling direct connection between bottom-up reaction networks and experimental observations. Highly efficient propagation of long-timescale kinetic profiles is required for automated reaction network refinement and is enabled here by a new discrete kinetic approximation.
  The resulting \textit{Kinetica.jl} simulation package therefore enables automated generation, characterization, and long-timescale modelling of complex chemical reaction systems. We demonstrate this for hydrocarbon pyrolysis simulated over timescales of seconds, using transient temperature profiles representing those of tubular flow reactor experiments. 
\end{abstract}

\newpage
\section{Introduction}

Many industrial components and materials are subjected to extreme environmental stresses from a number of sources, limiting their lifetimes. Heat generated through friction, oxidative conditions, and exposure to different wavelengths of electromagnetic radiation can all contribute to permanent material degradation at the molecular level, increasing the potential for critical material failure\cite{Petrova2020, Liu2012, Han2022, Celina2013, Tsotsis1995, Celina2019, Fillmore2004}. While these effects can be studied under controlled experimental conditions, this can involve many long-timescale trials. Even with such experimental results, it can be difficult to elucidate the microscopic mechanisms that lead to macroscopic degradation phenomena\cite{Porck2000}.

Alternatively, degradation processes can be modelled computationally. However, a deep understanding of all potential chemical reactions that a material \emph{could} undergo during experiments is required to fully capture the emergent chemical kinetics over long timescales. Not only does this understanding demand an expansive chemical reaction network (CRN) of elementary reactions, but also requires the rates at which these events occur (which are inherently dependent on the experimental conditions, such as temperature, at a given time)\cite{Broadbelt2005}.

This problem has been the focus of significant research for decades and many approaches for automated reaction discovery (ARD) have emerged for enumerating all possible reactions from a set of given chemical species\cite{Dewyer2018}. These can range from exhaustive searches through vast molecular graphs (sometimes referred to as connectivity or adjacency matrices) to molecular dynamics-based sampling techniques\cite{Gao2016, Wang2014}. Most modern approaches include a way to selectively explore chemical reaction space, only following reaction pathways that yield kinetically viable products under a given set of experimental conditions. This was pioneered by Susnow \latin{et. al.} in the \verb|NetGen| code, where estimated rates of formation of each unreacted species are iteratively used to expand a 'core' of species that can undergo reaction, growing the CRN only where reactions are predicted to be viable based on approximate rates of species formation\cite{Susnow1997}. This approach has the benefit of creating streamlined CRNs that are more easily analysed and simulated. It has seen use in popular contemporary codes such as the Reaction Mechanism Generator (RMG) and Software for Chemical Interaction Networks (SCINE)\cite{Gao2016, Bensberg2023}.

The determination of accurate reaction rates for the reactions sampled by ARD schemes is another point of difficulty when constructing large CRNs\cite{Dewyer2018}. In the standard approach, reaction rate evaluation requires identification of the transition state (TS) for every reaction in a CRN, typically starting from minimum energy path (MEP) schemes such as the nudged elastic band (NEB) method or the growing string (GS) method, which is coupled to a computationally expensive, but accurate, electronic structure method such as density functional theory (DFT)\cite{Jonsson1998, Peters2004}. There have been many recent developments that allow for the calculation or prediction of the rate constants within a CRN at reduced computational cost, but we will reserve this discussion for a forthcoming paper in which we investigate the applicability of using machine learning to predict reaction rates.

A further important challenge associated with simulations of degradation processes comes from the kinetic modelling of CRNs. This is the process of converting a CRN into a time-dependent form and propagating the degradation process in time under a set of given thermodynamic conditions. These forms can include discrete descriptions of every molecule being simulated, such as the chemical master equation (CME) and the stochastic simulation algorithms (SSAs) used to explicitly model every reactive event within a network\cite{Gillespie2007}. They can also include continuous, approximate descriptions such as the reaction rate equation (RRE), which is solved as a system of ordinary differential equations (ODEs)\cite{Higham2008}. Kinetic modelling of CRNs provides invaluable data, such as the distribution of all species in a reaction mixture at any given time. These distributions, as well as statistics like concentration fluxes, can be used to identify potentially hazardous byproducts of chemical degradation processes and avoid their formation\cite{Gopal2007}.

When modelling degradation over long timescales, however, there is an additional consideration. When simulating the kinetic evolution of a CRN under constant experimental conditions (temperature, pressure, etc.), assuming reaction rates respect the principle of detailed balance, the concentration of every species within the CRN will tend towards a steady state\cite{Makarov2017}. Real-world systems of interest are rarely held at constant conditions over extended periods. As such, a complete kinetic analysis of CRNs requires variable experimental conditions to tackle real-world, long-timescale kinetics. These variable conditions should be implemented to allow for arbitrary state variables to be utilised in kinetic simulations, such that simulations can be extended to take into account the effects of interesting and realisable changes in experimental conditions.

Kinetic simulations with variable experimental conditions are not often a priority of many contemporary CRN exploration codes, which instead focus on the generation and analysis of complex CRNs\cite{Zhao2021, Maeda2014, Jafari2018}, sometimes extending to constant condition kinetic simulations\cite{Gao2016}. On the other hand, codes such as Cantera and CHEMKIN exist for the sole purpose of performing complex kinetic and thermodynamic simulations on already explored networks\cite{cantera, Kee1996}. While these codes excel at performing reactive flow simulations under variable conditions, they are restricted to using a narrow set of ODE solvers that are not guaranteed to be capable of handling the solution of expansive, extremely stiff RREs over timescales many times greater than those of the individual reactions occurring within them. To tackle these problems reliably, we require access to a large library of ODE solvers that can be interchanged depending on the requirements of the kinetic simulation at hand.

\begin{figure}[h!]
    \centering
    \includegraphics{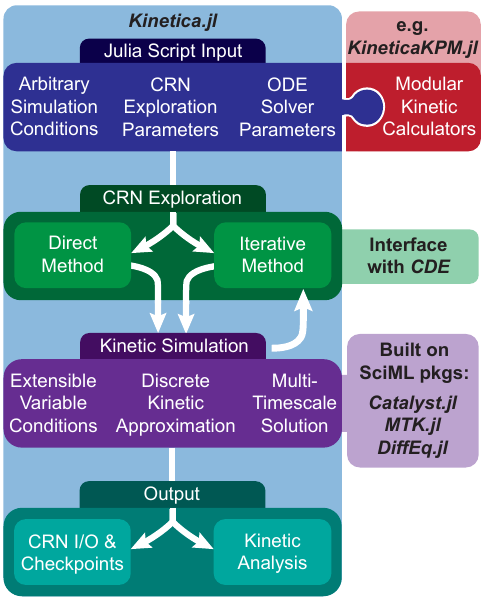}
    \caption{Schematic of the workflow within \textit{Kinetica.jl}, including a breakdown of major features and dependencies. Two dependencies, \textit{ModelingToolkit.jl} and \textit{DifferentialEquations.jl}, are abbreviated as \textit{MTK.jl} and \textit{DiffEq.jl} respectively.}
    \label{fig:Intro_Workflow}
\end{figure}

There is currently no unified code with which both CRNs for arbitrary variable experimental conditions can be automatically generated  \emph{and} long-timescale kinetic modelling under these conditions can be performed. Here, we present \textit{Kinetica.jl}, a new package written in the Julia language (Fig. \ref{fig:Intro_Workflow}). \textit{Kinetica} combines a kinetics-based reaction exploration algorithm with a molecular graph-driven sampling (GDS) approach to automatically explore the chemical reaction space that is relevant to arbitrary user-supplied variable experimental conditions. It makes use of an existing rich package ecosystem for symbolic-numeric computation and high-performance ODE solving to perform kinetic modelling of large CRNs for the long-timescale breakdown of molecules with continuously-variable conditions. The Julia language and its SciML organisation are vital to this approach, with packages such as \textit{DifferentialEquations.jl}, \textit{ModelingToolkit.jl} and \textit{Catalyst.jl} enabling fast simulations with simple low-level access to a plethora of underlying ODE solvers\cite{Bezanson2017, Rackauckas2017, Ma2021, Loman2022}. This flexibility allows us to extend the RRE solution with continuously variable conditions with an approximation that enables fast construction and simulation of large CRNs with negligible loss in accuracy.

This paper details the implementation of many of \textit{Kinetica}'s underlying systems and benchmarks some of the capabilities of the code on a generated CRN for the pyrolysis of ethane. However, \textit{Kinetica.jl} also contains a modular interface for kinetic calculators, separate packages which can be used to extend the available ways in which reaction kinetics can be modelled. We are releasing \textit{Kinetica.jl} alongside the \textit{KineticaKPM.jl} calculator for machine learning (ML)-based kinetic prediction, which uses a previously published activation energy prediction model to enable fast, computationally inexpensive CRN construction\cite{Ismail2022}.

The remainder of this article is organized to explain each of the major workflow sections shown in Fig. \ref{fig:Intro_Workflow}. In Section \ref{sec:MechanismExploration}, we describe two implementations of CRN exploration with our GDS approach, particularly focusing on the coupling between kinetic modelling and reaction exploration to provide an efficient scheme for automated CRN construction. In Section \ref{sec:kineticModelling}, we describe how \textit{Kinetica} performs long timescale kinetic simulations of generated CRNs under conditions of variable thermodynamic parameters, offering a route towards experimentally-tailored CRNs. Finally, in Section \ref{sec:CaseStudy} we showcase the end-to-end \textit{Kinetica} simulation workflow for ethane pyrolysis experiments. Finally, we offer some future directions for the further development of \textit{Kinetica}.

\section{Reaction Mechanism Exploration}\label{sec:MechanismExploration}

\textit{Kinetica} selectively explores chemical reaction space for molecular reactive processes under complex experimental conditions using a two-step approach. First, individual reactions and mechanisms are generated using a single-ended graph-driven sampling (GDS) approach, as developed in previous work\cite{Habershon2015}. Second, the generated reactions are aggregated into reaction networks that are selectively explored using an iterative kinetics-based algorithm, delivering extensive coverage of chemical reaction space while maintaining CRN compactness. In the following, the generation of CRNs in \textit{Kinetica} is discussed in the context of molecular degradation processes, but we emphasize that our overall approach is equally applicable to systems of multiple reactive species.

\subsection{Generating Reactions and Mechanisms}\label{sec:Exploration_mechanism_generation}

The GDS approach, referred to as the graph-based reaction-path sampling approach in refs. \citenum{Habershon2015,Ismail_2019,Habershon:2021aa,GDS_review,Habershon:2019ee}, was originally formulated for use in double-ended reaction path searches. Here, reactant and product Cartesian coordinates ($\mathbf{r}_0$ and $\mathbf{r}_P$, respectively) are transformed into connectivity matrices (CMs, also referred to as \emph{graphs} $\mathbf{G}^0$ and $\mathbf{G}^P$, respectively). The original double-ended GDS approach (DE-GDS) searches for directed paths through the chemical reaction space that connect the two end-point graphs; each such path represents a candidate reaction mechanism that may be post-analyzed using electronic structure calculations to evaluate and rank mechanistic hypotheses. 

In the context of GDS, the connectivity matrices $\mathbf{G}$ are $N \times N$ square matrices, where $N$ is the number of atoms in the system, with elements
\begin{equation}
    G_{ij}=\left\{\begin{array}{ll}
        1 & \text{if } r_{ij}<r_{ij}^{\text{cut}},\\ 
        0 & \text{otherwise.} 
    \end{array}\right.
\end{equation}
Here, $r_{ij}$ is the distance between atoms $i$ and $j$, and $r_{ij}^{\text{cut}}$ is a cutoff distance for atoms of the same types as $i$ and $j$, below which the atoms are considered to be bonded (noting that the \emph{type} of chemical bond is not relevant in this definition). 

Using a simulated annealing procedure, a sequence of CMs can be stochastically modified to generate potential mechanisms comprising elementary reactions such that reactant and product CMs, $\mathbf{G}^0$ and $\mathbf{G}^P$, are ensured to be connected. By geometry optimisation of the reaction-path intermediates, for example, using semi-empirical electronic structure methods, this approach delivers sequences of reactive intermediates, connected by elementary reactions, that form a low-energy path through chemical reaction space. Reaction paths generated in this way can subsequently be complemented with schemes to identify the MEP between each intermediate structure along the path. The double-ended GDS algorithm is implemented in the Chemistry Discovery Engine (CDE) code and has previously been used to investigate, for example, catalysis and reactions in the interstellar medium.\cite{Ismail_2019,Habershon:2019ee,Habershon:2021aa}

\begin{figure}[p]
    \centering
    \includegraphics{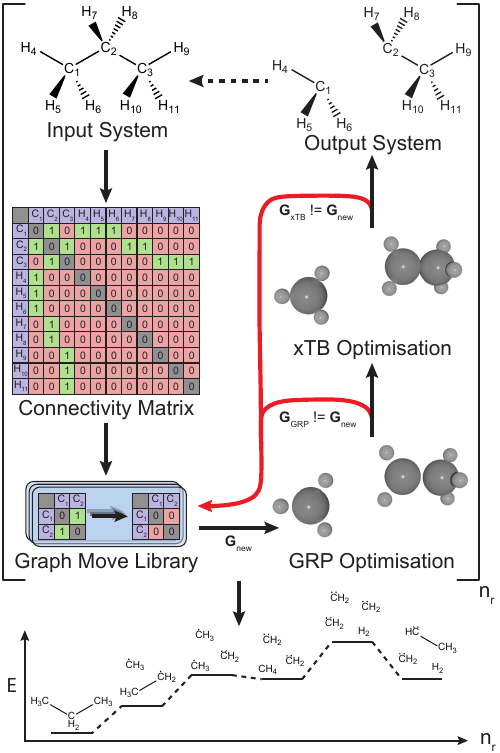}
    \caption{Schematic of the SE-GDS workflow within CDE\cite{Habershon2015,Ismail_2019,Habershon:2021aa,GDS_review,Habershon:2019ee}. A CM is extracted from an input system and a graph move is sampled from the graph move library. The move is applied to generate a new CM $\mathbf{G}_{\text{new}}$, which is converted back into a geometry by successive optimisation with the GRP and with GFN2-xTB\cite{Bannwarth2019}. If either changes the CM, a new graph move is sampled. The resulting geometries can be fed back in as inputs $n_r$ times to generate a random mechanism.}
    \label{fig:Exploration_CDE}
\end{figure}

While the DE-GDS method, by definition, yields reactions and intermediates that connect known reactants and products, the automated high-throughput generation of arbitrary CRNs instead demands a method that is suitable for generating reactions when there is no known end-point. We have therefore extended CDE with a single-ended GDS (SE-GDS) algorithm, the aim of which is to generate a CRN given only input knowledge of reactants. This uses many of the same principles as the double-ended approach but can be used without requiring knowledge of a final target structure. The SE-GDS algorithm is outlined in Fig. \ref{fig:Exploration_CDE}.

The SE-GDS algorithm begins by taking an input molecular geometry and calculating its CM. We then select a ``graph move'' to perform on this matrix. A graph move is defined as a specified change in the CM to a new state. Graph moves can be simple, such as making or breaking a bond between a randomly selected pair of atoms, or can be more complex, such as four-atom reactions typified by the dissociation of molecular hydrogen from across a carbon double-bond. Following our DE-GDS approach, mechanistic exploration uses a library of graph moves representing the allowed chemical reactivity; these moves can be tailored to explore particular types of chemical reaction, or can be simply left to be as broad as possible to promote wider exploration of novel chemical reaction paths. Furthermore, as well as supporting the specification of any number of allowed graph moves, our approach also allows users to define the elements that each graph move applies to, offering further routes to tailoring reaction exploration across the range from focused mechanistic hypothesis exploration to large-scale CRN generation. While the graph move library was used for stochastic selection of CM modifications in previous DE-GDS simulations, we have modified it here to also account for a relative probability of each graph move being selected during mechanism generation. This allows us to bias the selection of graph moves towards reaction types that are expected to be more or less prevalent in a given molecular system (again, noting that this bias can also be removed to drive unbiased CRN exploration).

Once a graph move (Fig. \ref{fig:Exploration_CDE}) has been randomly selected, a corresponding set of atoms is chosen to participate in the selected reaction. The graph move is then applied to the current CM of the system, and the resulting CM is verified to ensure that no modified atoms violate any user-defined valence or bonding constraints; for example, a typical constraint is to ensure that carbon atoms are bonded to no more than four other atoms, thereby respecting standard atomic valencies. The resulting CM is also checked against an (optional) list of forbidden graphs (or bonding patterns), which can be used to stop the formation of any unwanted or non-physical bonding patterns. 

If these checks are passed, an approximate molecular geometry for the system \emph{after} application of the reactive move is generated using a graph restraining potential (GRP); this is a simple empirical potential that is designed to be minimized if a particular set of atomic Cartesian coordinates $\mathbf{r}$ correspond to a target CM, $\mathbf{G}$. In other words, minimization of the GRP for target graph $\mathbf{G}$ with respect to the atomic coordinates $\mathbf{r}$ results in a structure for which the calculated CM matches $\mathbf{G}$. The GRP we use here is similar to that deployed in our previous studies,\cite{Habershon2015} and takes the following form:
\begin{equation}
    V\left( \mathbf{r},\mathbf{G} \right) = W\left( \mathbf{r},\mathbf{G} \right) + V_{\text{mol}}\left( \mathbf{r},\mathbf{G} \right).
\end{equation}
Here, $W\left( \mathbf{r},\mathbf{G} \right)$ is an atomic constraining potential that provides a force to move atomic coordinates $\mathbf{r}$ to match the connectivity in $\mathbf{G}$, and $V_{\text{mol}}\left( \mathbf{r},\mathbf{G} \right)$ is a molecular constraining potential which ensures that individual molecules are kept sufficiently far apart such that no bonds can form between their constituent atoms. 

The GRP allows us to quickly create approximate geometries of newly sampled molecular products from their CMs only. Once new structures have been successfully generated by optimization under the GRP, the resulting (typically high-energy) geometries can then be optimised to stable geometries using an electronic structure method. If any of the valence-based validity checks fail after GRP and electronic structure-based optimization, or if the CM proposed by the graph move is modified by either of these optimisations, the CM is restored to its original form (before the new reaction was imposed) and a new graph move is generated until either a valid graph move is identified or a maximum number of attempts have been made (suggesting that the graph move library cannot be successfully applied to any sub-graph of the current CM).

This process of CM generation, graph move selection/validation and geometry optimisation forms one iteration of the single-ended mechanism search, which can be repeated for a user-specified number of iterations $n_{r}$ to build up a randomly-sampled mechanism of reactions with optimised intermediates from a given starting system. Finally, this whole workflow can be repeated $n_{m}$ times to generate multiple mechanisms (each comprising $n_{r}$ reactions).

The importance of geometry relaxation of GDS intermediates cannot be overstated, as it effectively sanitises GRP predictions. Because the GRP is biased to generate a geometry that respects the new CM for the system, it can occasionally lead to unstable molecular geometries that isomerise when optimised on a more accurate potential energy surface (PES). For this reason, we employ the semi-empirical GFN2-xTB method\cite{Bannwarth2019} to perform the secondary optimisations within the SE-GDS, as it has been shown that it often produces atomic forces and optimised geometries similar to DFT within small-to-medium-sized molecules while also exhibiting excellent computational efficiency. Given that a single run of the SE-GDS algorithm deployed here typically requires more than $2n_{r}n_{m}$ secondary optimisations to verify GRP accuracy, the GFN2-xTB method represents a good balance between accuracy and computational cost. As a representative example, a SE-GDS run with a 15-mer of polyethylene with $n_{r}=400$ and $n_{m}=1$ yields a mechanism (i.e. set of GRP- and PES-optimized intermediate structures) in a matter of minutes on a modern consumer-grade computer.

\subsection{Network Assembly}\label{sec:networkAssembly}

\begin{figure}
    \centering
    \includegraphics{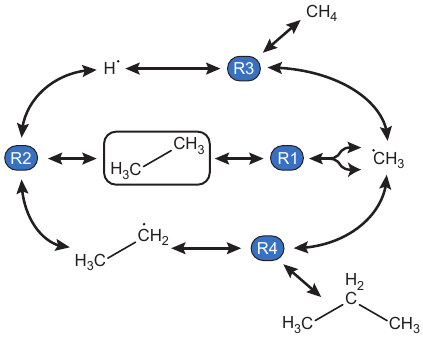}
    \caption{Example of a potentially infinite CRN. Free radical species, formed by homolytic cleavage of bonds in ethane, can combine to form hydrocarbons larger than the original reactant --- in R4 an ethyl radical can form a bond with a methyl radical to become propane. This process can be repeated to continue forming larger hydrocarbon chains.}
    \label{fig:Exploration_Ethane_initial_CRN}
\end{figure}

With a method for generating possible reactions in hand, it is possible to think of many ways in which a reaction network could be assembled. However, the stochastic nature of the SE-GDS process means that the way in which chemical reaction space is sampled plays an important role in how complete (or ``correct'') a generated CRN is in capturing the reactions important to long-timescale chemical kinetics. Furthermore, under some simulation conditions, CRNs have the potential to grow indefinitely --- unless constraints on system size are imposed --- as demonstrated by molecular degradation into free radical species, followed by radical chain growth (Fig. \ref{fig:Exploration_Ethane_initial_CRN}). Care must therefore be taken to sample chemical reaction space in a way that captures only the chemical reactions that are possible under a set of experimental conditions, rather than seeking to address the combinatorially-difficult task of generating \emph{every} possible reaction.

In the next two sections, we discuss two routes to CRN construction using SE-GDS. First, we present a ``direct'' or brute-force approach; as discussed below, this route quickly runs foul of the challenges associated with the complexity and size of available chemical reaction space. Second, we present an iterative scheme for CRN construction using SE-GDS; this iterative approach is based on using chemical kinetics simulations to grow a CRN along reaction pathways that are the most ``kinetically viable''. As we show later, this iterative approach results in focused CRNs that better reflect the essential kinetics of complex variable-condition degradation processes.

\subsubsection{Direct Exploration}\label{sec:networkAssembly_direct}

\begin{figure*}[t]
    \centering
    \includegraphics{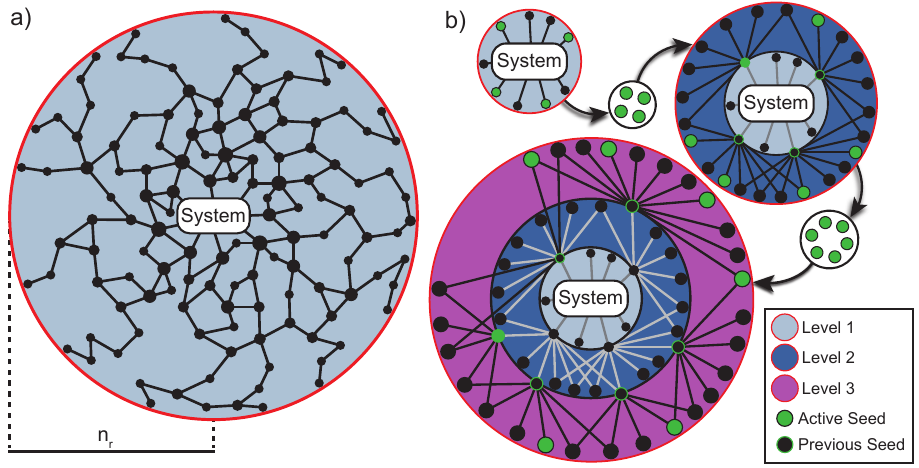}
    \caption{Schematics of the two CRN exploration methods provided in \textit{Kinetica}. a) The direct method uses SE-GDS to generate random mechanisms of length $n_r$ from a starting system of molecules until the CRN is converged. Stochastic sampling leads to under-converged species on the outer edge of the explored reaction space. b) The iterative method proceeds in levels, with each level sampling the reactions adjacent to ($n_r = 1$) the high concentration species (seeds) of the previous level and running a kinetic simulation to determine the seeds for the next level, avoiding under-converged species above a threshold.}
    \label{fig:Exploration_Methods}
\end{figure*}

The simplest method of building a CRN for chemical breakdown reactions is to use SE-GDS to sample as many mechanisms as possible, each for a defined number of reactions $n_r$ (Fig. \ref{fig:Exploration_Methods}a). Here, $n_r$ can be thought of as an effective ``radius'' that defines the outer bounds of explored reaction space. By sequentially adding newly-discovered species to the growing network and connecting them with the sampled reactions, vast CRNs can easily be constructed. We typically ignore reactions of a higher molecularity than bimolecular reactions because their probability of occurrence in reaction mixtures is negligible \cite{Laidler1963}, although this is easily modified.

This method is perfectly valid for small CRNs where every possible species is only a few reactions away from the starting species ($n_{r} \leq 10$); in such cases, the probability of repeatedly generating intermediate species --- and their possible reactions --- is relatively high. In this way, even by randomly sampling new reactions, the network is likely to converge (such that all possible reactions and species allowed by a given set of graph moves will have been found). This convergence can be determined by repeatedly sampling new mechanisms until no new reactions have been found within a user-defined number of attempts.

However, as $n_{r}$ is increased to grow larger CRNs, two problems begin to emerge in this direct method. First, CRNs explored this way should sample every possible reaction between every pair of species, leading to a combinatorial growth of reactions as more species are discovered, which can quickly make kinetic simulations of the resulting CRNs computationally intractable. Second, the combinatorial explosion of reactions results in the probability of completely sampling a given species (and all of its reactions) becoming smaller for species that are ``further'' from the initial reactive species. For example, if a species can only be created by one specific mechanism consisting of many reactions, then completely sampling the reactions of this species relies on being able to repeatedly follow this mechanism in many randomly-driven SE-GDS runs. In an ideal scenario with infinite exploration time, all reactions could be found stochastically; with finite exploration time and a user-defined convergence cutoff, the CRN can instead become more sparse as the exploration proceeds away from the initial species. In other words, many species may exist on the outermost discovered ``edge'' of explored reaction space, and their reactions with both discovered and undiscovered species may be under-sampled.

Under-sampled species are not necessarily a problem if their formation is not kinetically viable --- if the concentration of such species in the kinetic simulation of the resulting CRN is always low, then there is no need to sample the further onward reactivity of these species. However, if these under-sampled species are formed in high concentrations during kinetic modelling, they can cause one of two effects. If the reverse reactions leading to decomposition of the under-sampled species have very high rate constants (i.e. they are barrierless, or have very low energetic barriers to reaction), then the species is likely to be so short-lived that any contributions to the kinetics of the CRN from further additional reactions stemming from it are negligible. In this case, the convergence of the CRN would likely not be impacted by the under-sampled species. If, however, there are high activation barriers to these reverse reactions of the under-sampled species (likely if, for example, the formation of the under-sampled species is highly exothermic), then it is likely to have a lifetime within the reaction mixture that supports further reaction. Within the framework of this direct approach to CRN construction, completely sampling these species is only possible by increasing the convergence criteria and waiting for an unlikely chain of reactions leading to this species to be found again, so that further reactions can be discovered for the under-sampled species.

As we show later, these problems are significant enough to dramatically influence the emergent kinetics for CRNs sampled using the direct method in combination with SE-GDS. As such, in the following, we propose an alternative iterative approach for large-scale CRN construction and simulation.

\subsubsection{Iterative Exploration}\label{sec:networkAssembly_iterative}

\textit{Kinetica} implements an iterative exploration algorithm that uses the results of a kinetic simulation at each iteration to identify high-concentration species, the reactions of which are explored and added to the growing CRN in the next iteration. This algorithm (Fig. \ref{fig:Exploration_Methods}b) begins by using SE-GDS to explore a sub-space of reactions with species adjacent to the starting species (where adjacent species are those that are a single elementary reaction away, so $n_r=1$), until convergence is deemed to have been reached in the same way as was presented in the direct exploration method. These reactions form the initial \emph{level} of the overall CRN. This level is then kinetically modeled at the experimental conditions of interest, and all species with maximum concentration over the entire simulation ($c_{\text{max}}$) greater than a user-defined concentration cutoff ($c_{\text{select}}$), are marked as ``seed'' species. These seed species are collated into a single seed system, which is then passed back to SE-GDS as starting structures for the next level of CRN exploration.

Within subsequent levels of exploration, reactions adjacent to the input seed system are sampled until level convergence, and a new kinetic simulation is performed using the combined CRN created by adding the reactions and species explored in the current level to those from previously explored levels. The seed species for the next iteration are again identified and collated into a new seed system. Seeds can be selected from anywhere in the currently explored CRN, meaning it is common to see seeds persisting throughout multiple levels of exploration when reactions have not been found that significantly reduce their maximum concentrations. Similarly, if new reactions are discovered that re-form a previous seed, such a seed can also re-enter the current seed system and be exposed to reactions with other more recently discovered species.

The concentration cutoff $c_{\text{select}}$ must be chosen carefully however, as setting this cutoff too high selects too few seed species at each iteration, leading to an incomplete network. On the other hand, setting the cutoff too low would select too many seed species at each iteration, resulting in excessive numbers of levels being explored (at high computational cost). Only species with final concentrations above $c_{\text{select}}$ should therefore be considered accurate, as molecules that have not been selected as seeds can be potentially under-sampled. In practice, we find that is is best to run multiple iterative explorations with decreasing values of $c_{\text{select}}$, such that an acceptable tradeoff can be established between CRN complexity and accuracy.

Iterative CRN growth continues until the CRN is converged --- that is, all reactions and species relevant to the reactivity of the initial species under a set of experimental conditions have been identified. Unlike the convergence criteria used in the direct method or level convergence, iterative CRN convergence is indicated when no further changes occur across a user-defined number of consecutive seed systems, indicating that the current seeds represent the high-concentration species that will be present in the full CRN. This iterative method has the added benefit of selectively exploring chemical reaction space and only following kinetically viable paths, negating much of the effect of the combinatorial growth seen in the direct exploration method and leading to fewer reactions required in the final CRN, in turn allowing for more efficient kinetic simulations (see Section \ref{sec:kineticModelling}). 

\subsubsection{Comparison to Contemporary Methods}

As noted previously, kinetically-driven reaction exploration algorithms like the above have already seen extensive use in popular software codes such as NetGen and RMG\cite{Susnow1997, Gao2016}. These codes differentiate between 'reacted' species (those inside a seed system) and 'unreacted' species (those outside of a seed system) by using estimated reaction flux to determine the species that are of kinetic importance. At each iteration, unreacted species with reaction flux above a user-defined cutoff are added to a 'core' of reacted species, for which all possible reactions are subsequently generated by matching functional groups within reactants to reaction templates that can be applied in a similar way to the graph moves of SE-GDS to generate product structures.

However, these codes are limited to performing CRN construction under a fixed set of experimental conditions. The CRN exploration algorithms they implement would not be applicable for the long-timescale simulations under variable conditions that we wish to achieve. Because reacted species are never removed from the reactive core in this algorithm, the number of new species pairs that must be considered at every iteration, $p_{\text{new}}$, is:
\begin{equation}
    p_{\text{new}} = s_{\text{new}} \left( s_{\text{prev}} + 1 \right),
\end{equation}
\noindent where $s_{\text{prev}}$ is the number of species in the previous iteration's core, and $s_{\text{new}}$ is the number of species being added by the current iteration. Since each species pair can produce many reactions, the sampling process in each iteration becomes progressively more expensive as the core grows. This is less of a problem in static kinetic simulations, especially if the core remains relatively small. However, in variable kinetic simulations, the continuously changing rate constants of reactions could yield high concentration fluxes over the timespan of each kinetic simulation, leading to many of the unreacted species in each iteration being added to the core. Since both $s_{\text{prev}}$ and $s_{\text{new}}$ would therefore be large in each iteration, many new species pairs (each with many new reactions) would have to be considered in the next iteration.

In contrast, we anticipate that our iterative approach, using species' \emph{maximum} concentrations and refreshing the core (i.e. seed system) at each iteration, will result in a much smaller subset of reactions added in each iteration. This additional granularity is vital for achieving efficient CRNs that can be kinetically modelled over long timescales under challenging experimental conditions, as we demonstrate below.

\section{Kinetic Modelling}\label{sec:kineticModelling}

To perform kinetic modelling of a CRN, generated by the SE-GDS-based approaches described in Section \ref{sec:MechanismExploration}, it must be transformed into a form that can be numerically integrated with time. When considering simulations of large CRNs in the continuous concentration domain, this typically involves transforming a network into a reaction rate equation (RRE), which acts as a deterministic approximation of the underlying stochastic chemical master equation (CME)\cite{Gillespie2007}. In the CME, each reaction is stochastically sampled and resolved sequentially by directly updating the number of molecules of the affected species in the reaction mixture. In the RRE, all reactions are considered to be occurring simultaneously and each species has a continuous concentration gradient which dictates how its concentration evolves at future points in time.

RREs are typically formulated as a set of coupled ordinary differential equations (ODEs). Consider the following toy CRN:
\begin{align*}
    \ch{
        A & <=>[ k_{1} ][ k_{-1} ] B + C \\
        B & <=>[ k_{2} ][ k_{-2} ] D \\ 
        C + D & <=>[ k_{3} ][ k_{-3} ] E,
    }
    \label{eqn:ToyNetwork}
\end{align*}
\noindent where $A$, $B$, $C$, $D$ and $E$ represent five chemical species, and $k_{1}$, $k_{2}$, $k_{3}$, $k_{-1}$, $k_{-2}$ and $k_{-3}$ represent the mass action rate constants\cite{Erdi1989} for three forwards and backwards reactions, respectively. The resulting RRE for this system would be:
\begin{equation}
\begin{split}
    \frac{\mathrm{d} A\left( t \right)}{\mathrm{d}t} & = - k_1 A\left( t \right) + k_{- 1} B\left( t \right) C\left( t \right) \\
    \frac{\mathrm{d} B\left( t \right)}{\mathrm{d}t} & = k_1 A\left( t \right) + k_{- 2} D\left( t \right) - k_2 B\left( t \right) - k_{- 1} B\left( t \right) C\left( t \right) \\
    \frac{\mathrm{d} C\left( t \right)}{\mathrm{d}t} & = k_1 A\left( t \right) + k_{- 3} E\left( t \right) - k_{- 1} B\left( t \right) C\left( t \right) - k_3 C\left( t \right) D\left( t \right) \\
    \frac{\mathrm{d} D\left( t \right)}{\mathrm{d}t} & = k_2 B\left( t \right) + k_{- 3} E\left( t \right) - k_{- 2} D\left( t \right) - k_3 C\left( t \right) D\left( t \right) \\
    \frac{\mathrm{d} E\left( t \right)}{\mathrm{d}t} & = - k_{- 3} E\left( t \right) + k_3 C\left( t \right) D\left( t \right).
\end{split}
\label{eqn:ToyNetworkStatic}
\end{equation}

The set of RREs above are trivial to solve with many modern computational ODE solvers. However, as the number of species in a CRN increases, the number of reactions connecting them is expected to increase dramatically. This results in many more terms per ODE in the resulting system. Additionally, rate constants in large CRNs often span many orders of magnitude, yielding complex and stiff systems of ODEs that are non-trivial to propagate.

\textit{Kinetica.jl} makes heavy use of packages in the Julia language's SciML organisation to perform these kinetic  simulations\cite{Rackauckas2017}. We begin by composing sampled reactions together using \textit{Catalyst.jl}, a symbolic modelling package which allows us to programatically construct and analyse arbitrarily-sized CRNs\cite{Loman2022}. The resulting \verb|ReactionSystem| objects are passed to \textit{ModelingToolkit.jl}, a package for performing high-performance symbolic-numeric computation\cite{Ma2021}. Within \textit{ModelingToolkit.jl}, we can compile a \verb|ReactionSystem| into a symbolic \verb|ODESystem|, where the time-dependent species concentrations are represented by symbolic variables (runtime-determined continuous values), and the rate constants that control their evolution are represented by symbolic parameters (runtime-defined static values). The resulting functions are capable of high-performance computation, and the symbolic nature of this compilation allows the same basic system of ODEs to be used for many calculations with varying parametric rate constants.

Alongside the direct compilation of the symbolic RRE, \textit{ModelingToolkit.jl} can be used to perform a variety of performance optimisations. Within \textit{Kinetica.jl}, \textit{ModelingToolkit.jl} performs automatic sparsity detection such that sparse linear algebra routines can be used to more efficiently integrate the RRE\cite{Gowda2019}. Having the full functional form of the RRE also enables \textit{ModelingToolkit.jl} to automatically compile an analytic expression for the Jacobian of the \verb|ODESystem|. These performance optimisations are vital for performing long-timescale calculations in an efficient and timely manner.

Once generated, the symbolic RREs are solved by ODE solvers implemented within \textit{DifferentialEquations.jl}, an extensive collection of numerical differential equations solvers implemented both natively within Julia, and also within other high-performance languages such as C and Fortran, packaged together under a common interface\cite{Rackauckas2017}. With this library, the selected ODE solver can be changed with a single line of code, enabling the highest degree of flexibility when solving numerically stiff RREs. \textit{DifferentialEquations.jl} allows binding of numeric values to the symbolic parameters created by \textit{ModelingToolkit.jl}, defining initial conditions (the starting concentration of each species) and the simulation timespan.

With this workflow, it is relatively straightforward to convert an explored CRN into high-performance code that can be numerically integrated to reveal concentration-time profiles for all species in the network. However, because each reaction's rate constant is implemented as a runtime-defined parameter, these values must be provided for the simulation to commence.

\subsection{Kinetic Calculators}

To calculate individual reaction rate constants, \textit{Kinetica.jl} includes several kinetic calculators; these represent a key element of extensibility within \textit{Kinetica} that enables connection to experimental observables. Each calculator defines the experimental conditions (e.g. temperature, pressure, volume, etc.) that it can accept; users can symbolically pass the values of these conditions at simulation time to calculate rate constants on-the-fly. Crucially, the structure of these calculators is simple to implement, and users can define their own calculators that accept custom conditions to run simulations with custom rate expressions.

\textit{Kinetica.jl} provides a base calculator, \verb|PrecalculatedArrheniusCalculator|, as an example implementation (see Supplementary Section S1.1 for example code). This calculator uses the Arrhenius equation to calculate a rate constant for each reaction at the provided temperature:
\begin{equation}
    k = A \exp \left[ -\frac{E_a}{RT} \right].
\end{equation}
It therefore takes a vector of $N$ activation energies, a vector of $N$ Arrhenius prefactors, and a named argument \verb|T| (representing the temperature of the reaction mixture) and uses them to calculate $N$ reaction rate constants, where $N$ is the number of reactions in the CRN. 

In the case of the iterative CRN exploration algorithm described in section \ref{sec:networkAssembly_iterative}, the set of reactions --- and hence the set of values for $E_a$ and $A$ for every reaction --- are not known \latin{a priori}. It is therefore important to also have calculators that enable determination of rate constants on-the-fly --- alongside network exploration. We provide one such calculator in the \textit{KineticaKPM.jl} package, which is capable of quickly estimating reaction rate constants using machine-learned values of $E_a$ for each reaction, as described in a previous work\cite{Ismail2022}. Other calculators could easily be envisioned for performing different roles - for example, a DFT-driven calculator for the evaluation of \latin{ab initio} rate constants using transition state theory could be easily implemented.

The kinetic calculators described can be called at the start of a simulation, with user-provided fixed conditions, to assign values to the symbolic rate constants in generated RREs; however, this fixes the values of any reaction rate constants for the duration of the kinetic simulation. This simulation technique represents the `standard' kinetic analysis of CRNs but is not applicable for simulating CRNs under variable experimental conditions.

\subsection{Simulations with Continuous Variable Conditions}

To achieve CRN kinetic simulations under variable conditions, \textit{Kinetica} includes a library of parametric condition profiles. These profiles can be flexibly defined as direct functions of time, or as gradient functions that are integrated in time to obtain the final condition profile. All profiles are agnostic to the quantity that they represent and can be bound to a symbolic parameter that allows for their value at any time within a simulation to be passed to the current kinetic calculator. This approach therefore enables kinetic simulations of CRNs to be performed using variable external conditions, providing a clear connection between computation and experiment.

Individual condition profiles are bound together within a \verb|ConditionSet|. Each \verb|ConditionSet| can be composed of a combination of condition types: 
\begin{enumerate}[label=\emph{\alph*})]
    \item fixed conditions, where the value of a condition is held constant for the duration of a simulation,
    \item directly variable conditions, where the variation of a condition with time can be expressed analytically,
    \item gradient-variable conditions, where the variation of a condition with time must be integrated from an analytically-expressible gradient function.
\end{enumerate}
These conditions are then passed into one of \textit{Kinetica}'s solvers, where the different condition types can be correctly employed within the RRE propagation. For example, if a simulation was desired in which the simulation temperature decreased, the pressure increased and the volume was held constant, this could be implemented with the \verb|ConditionSet| shown in Figure \ref{code:ConditionSet_example}.

\begin{figure}
    \centering
    \begin{minted}[
        baselinestretch=1.1,
        breaklines,
        linenos,
        numbersep=3pt,
        frame=lines,
        fontsize=\footnotesize,
        framesep=2mm
    ]
    {julia}
conditions = ConditionSet(Dict(
    :T => LinearDirectProfile(;
        rate = -10.0,
        X_start = 400.0,
        X_end = 285.0
    ),
    :P => LinearGradientProfile(;
        rate = 40.0,
        X_start = 1e5,
        X_end = 1.1e5
    ),
    :V => 1000.0
))
    \end{minted}
    \caption{Example \texttt{ConditionSet} definition featuring directly variable temperature, gradient-variable pressure and constant volume.}
    \label{code:ConditionSet_example}
\end{figure}

\noindent Here a directly variable temperature profile is used alongside a gradient-variable pressure profile for the sake of demonstration; for simple linear condition profiles the two can be used interchangeably, but more complex profiles may lend themselves more naturally to a gradient-based definition. The temperature profile generates the following function of simulation time:
\begin{equation}
    T\left( t \right) = 
    \begin{cases}
        \texttt{X\_start}, & \text{if } t \leq 0.0 \\
        \texttt{X\_start} + t\left( \texttt{rate} \right), & \text{if } t > 0.0 \text{ and } t \leq t_{\text{end}} \\
        \texttt{X\_end}, & \text{if } t > t_{\text{end}} \\
    \end{cases}
    \label{eqn:Tprofile_linear}
\end{equation}
\noindent where $t_{\text{end}} = \left( \texttt{X\_end} - \texttt{X\_start} \right) / \texttt{rate}$. The pressure profile generates the following function of simulation time:
\begin{equation}
    \frac{\mathrm{d} P\left( t \right)}{\mathrm{d}t} = 
    \begin{cases}
        0.0, & \text{if } t \leq 0.0 \\
        \texttt{rate}, & \text{if } t > 0.0 \text{ and } t \leq t_{\text{end}} \\
        0.0, & \text{if } t > t_{\text{end}} \\
    \end{cases}
    \label{eqn:Pprofile_linear}
\end{equation}
\noindent where $t_{\text{end}}$ is defined the same as above. 

\subsubsection{Implementation of Variable-Condition Kinetics}\label{sec:ContinuousVariableKinetics}

We achieve kinetic simulations under variable experimental conditions by incorporating condition profiles like the ones above into the RREs that are generated from CRNs with \textit{Catalyst.jl} and \textit{ModelingToolkit.jl}. This is accomplished within a secondary \textit{ModelingToolkit.jl} \verb|ODESystem|, where condition profiles are coupled to symbolic variables representing their values (or the values of their gradients, for gradient-based profiles) at a given simulation time. Within this \verb|ODESystem|, the symbolic rate constants are also bound to the output of the kinetic calculator, when called with the symbolic variable conditions as function arguments. Together, this forms a system of differential algebraic equations (DAEs) which can be integrated in time to calculate the rate constants of the reactions in the current CRN as they vary with the conditions provided in the input \verb|ConditionSet|.

\begin{figure*}[t]
    \centering
    \includegraphics{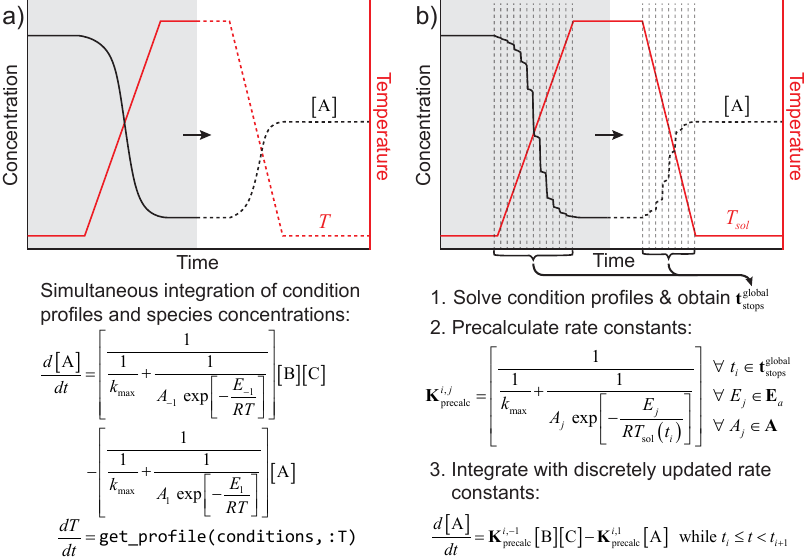}
    \caption{Differences between the two variable-rate kinetic simulation methods in \textit{Kinetica}. a) The continuous rate update formalism embeds the functional form of the rate constant of every reaction into the compiled RRE, allowing reaction rates to evolve continuously with experimental conditions and time. b) The discrete rate update formalism pre-calculates rate constants at discrete time points using interpolated solutions of condition profiles --- here $T_{\text{sol}}$ represents a solution of a temperature profile. During a kinetic simulation, the rate constants are updated at these time points to simplify RRE compilation.}
    \label{fig:Modelling_continuous_vs_discrete}
\end{figure*}

Combining these equations with the generated RRE forms another system of DAEs. DAEs are less easily solved by numerical methods than ODEs, so we use ModelingToolkit to automatically perform a set of substitutions that turn the DAEs back into ODEs. In this case, the substitution is simple - each rate constant equation (defined within the selected kinetic calculator) gets directly substituted into the relevant locations inside the RRE. Using a \verb|ConditionSet| with only variable temperature and the \verb|PrecalculatedArrheniusCalculator|, the first ODE in our toy CRN above therefore becomes:
\begin{equation}
\begin{split}
    \frac{\mathrm{d} A\left( t \right)}{\mathrm{d}t} = & - k_{1}\left( T \right) A\left( t \right) + k_{- 1}\left( T \right) B\left( t \right) C\left( t \right) \\
    = & A_{1} \exp\left [ -\frac{E_{a, 1}}{k_{B}T\left(t\right)} \right ] A\left( t \right) \\
    & + A_{-1} \exp\left [ -\frac{E_{a, -1}}{k_{B}T\left(t\right)} \right ] B\left( t \right) C\left( t \right).
\end{split}
\label{eqn:ToyNetworkContinuousA}
\end{equation}
\noindent Note that the resulting ODE is purely a function of time. We perform this substitution for the entire RRE to obtain a fully condition-dependent system of ODEs, which describes the evolution of the CRN with time and all the continuously variable conditions in the \verb|ConditionSet| (Figure \ref{fig:Modelling_continuous_vs_discrete}a).

\subsubsection{Performance Considerations}

Solving stiff systems of ODEs, such as those created by complex CRNs, is an inherently multiscale problem. Some reactions, including many bond-breaking events, have large activation energies, leading to very small rate constants and very slow reactions. Other reactions, such as radical recombinations, can be barrierless processes which occur as frequently as the collisions between reactants.

\textit{DifferentialEquations.jl} contains implementations of many solvers across a number of programming languages, so the choice of solver is key to achieving efficient kinetic simulations. We recommend using the \texttt{CVODE\_BDF} solver from the \textit{Sundials} suite\cite{Hindmarsh2005}. It is an implicit adaptive-order multistep VODE (variable-coefficient ODE) solver that can handle both large ODE systems and very high stiffness, and it has performed best for the systems considered here.

In recent years, there has been significant research into performing high-efficiency calculations on stiff systems such as CRN RREs by using methods other than implicit VODE solvers\cite{Verwer1995, Morii2020}. Indeed, simulations using implicit VODE solvers can suffer from high computational expense due to the complexity of numerical Jacobian evaluation and decomposition in each timestep.

Most of these methods focus on ways of partitioning the full CRN into groups of reactions with different timescales, effectively removing the stiffness from the ODE system. This is the foundation of methods such as the quasi-steady state approximation\cite{Schauer1983}, computational singular perturbation (CSP)\cite{Ardema1990} and the intrinsic low dimensional manifold (ILDM)\cite{Maas1992}, which utilise the Jacobian of the system to identify groups of reactions that can be separated by timescale. This, however, would be unfeasible for variable conditions --- species concentrations and rate constants are both time-dependent, so the Jacobian of the system must be diagonalised often to determine which reactions should be separated. This can be viable for small CRNs where these operations are relatively inexpensive, but repeated diagonalisation of large Jacobians would incur significant computational expense.

Alternate methods exist for instead estimating the characteristic timescales of reactions by less expensive means, making explicit multi-timescale solutions of large CRNs computationally cheaper to obtain than their respective implicit VODE solutions\cite{Gou2010}. However, since we symbolically model the CRN within \textit{ModelingToolkit.jl} and can therefore construct analytic Jacobians that take advantage of sparse linear algebra routines, we can greatly increase ODE solution speed without needing to perform any explicit CRN separation\cite{Perini2014}.

Despite the efficiency of the ODE solution algorithms in \textit{Kinetica}, the extremes of both fast and slow reaction types can still cause problems during numerical integration. We therefore implement several additional performance optimisations, as follows:

\paragraph{Handling Slow Reactions:}

Some reactions proceed so slowly that, even at high reactant concentrations, the resulting product concentrations will be negligible over the timescale of the simulation. These reactions can simply be removed from the CRN to maximise efficiency of the \verb|ODESystem| compilation.

We assume a reaction $i$ can be removed from the CRN if, over the entire timescale of the simulation, the maximum rate of that reaction $r_i^{\text{max}}$ is smaller than the precision, $rtol$, of the ODE solver:
\begin{gather}
    \textup{remove if } t_{\text{global}}\cdot r_{i}^{\text{max}} < rtol \\
    \textup{where } r_{i}^{\text{max}} = c_{\text{max}}^{2}k_{i}^{\text{max}} \nonumber
    \label{eqn:LowRateCutoff}
\end{gather}
\noindent where $t_{\text{global}}$ is the global simulation time (see below), $k_{i}^{\text{max}}$ is the rate constant for the reaction at the maximum temperature that will be seen within a simulation, and $c_{\text{max}}$ is a maximum concentration. This concentration should be a value that is greater than the maximum concentration any species in the reaction mixture should obtain. We recommend setting this at 2--5 times the concentration of the initial reactants. This ensures that any reactions that are removed from the CRN would never be formed in quantities greater than the numerical noise in the simulation, below which results should not be considered to be accurate.

\paragraph{Handling Fast Reactions:}\label{sec:ContinuousKinetics_FastReactions}

Very fast reactions are more problematic to kinetic propagation. The faster a reaction proceeds, the smaller the required timestep of the ODE solver. If the timestep is very small, many timesteps have to be taken to numerically integrate the system. In some cases, fast reactions may demand that the timestep falls to such a small value that it is less than the smallest precision $\epsilon$ that is computationally representable by the global simulation time $t_{\text{global}}$. This causes floating point underflow to occur when this timestep is added to the simulation time, resulting in the simulation time not advancing. 

We typically represent $t_{\text{global}}$ with an IEEE 64-bit floating point number (referred to hereafter as a \verb|Float64|) where $\epsilon = 2^{-53} \simeq 10^{-16}$. The smallest value that can be added to such a number is therefore $t_{\text{global}}\epsilon$. Simulation time accumulation therefore fails when
\begin{enumerate}[label=\emph{\alph*})]
    \item $t_{\text{global}}$ is so large that the timestep being added to it is less than $t_{\text{global}}\epsilon$,
    \item the simulation conditions are so favourable to reactions that the timestep must fall below $t_{\text{global}}\epsilon$ to resolve fast reactions,
    \item a combination of the above occurs.
\end{enumerate}
\noindent This problem is therefore intrinsic to long-timescale kinetic simulations. While calculations with numeric types of greater precision are possible, very few sparse matrix solvers are designed to perform operations on numeric types other than \verb|Float64|, and those that can handle arbitrary numeric types are currently too inefficient to enable adequate performance.

\textit{Kinetica} tackles the problem of fast reactions on two fronts. First, we allow for limiting the maximum rate constant. In a fluid, the maximum possible frequency of collision is the limited by the diffusion rate. We can therefore use partial diffusion control to place a hard limit on the maximum rate\cite{Rabinowitch1937}. In diffusion-controlled reactions, the rate constant $k_{D}$ is
\begin{equation}
    k_{D} = \frac{8RT}{3\eta}
    \label{eqn:DiffControlRate}
\end{equation}
\noindent where $\eta$ is the viscosity of the fluid at temperature $T$. Partial diffusion control limits the total rate of reaction $k_{r}$ as follows:
\begin{equation}
    k_{r, i} = \dfrac{1}{\dfrac{1}{k_{D}} + \dfrac{1}{k_{i}}}
    \label{eqn:PartialDiffControlRate}
\end{equation}
\noindent This ensures a smooth transition between the calculated rate constants and diffusion control.

However, even with a maximum rate constant in place, some combinations of variable conditions and overall timescale can still lead to floating point underflow during propagation. We sidestep this issue with a multi-timescale approach to simulation time accumulation. In this approach, a single simulation can be split up into chunks of simulation time, each of length $\tau_{c}$. When a kinetic simulation is initialized with species concentrations $\mathbf{c}=\mathbf{c_{0}}$, global simulation time $t_{\text{global}}=0.0$, local (chunk) simulation time $t_{\text{local}}=0.0$ and number of chunks $n_{c}=0$, it is only allowed to proceed until $t_{\text{local}}=\tau_{c}$, at which point the simulation pauses. The solver is then reinitialised at $t_{\text{local}}=0.0$ with $\mathbf{c}=\mathbf{c_{\text{final}}}$, where $\mathbf{c_{\text{final}}}$ is the vector of concentrations at the end of the previous chunk, and $n_c$ is incremented. 

This approach is continued with
\begin{equation}
    t_{\text{global}}=t_{\text{local}}+\tau_{c}n_{c}
    \label{eqn:tGlobal}
\end{equation}
\noindent until $t_{\text{global}}$ reaches the end of the global simulation time $t_{\text{end}}$. By replacing $t_{\text{global}}$ with the smaller $t_{\text{local}}$ during timestep accumulation, we allow small timesteps to be taken down to the value of $t_{\text{local}}\epsilon$. This lets us arbitrarily extend the numeric precision of simulations, ensuring we avoid floating point underflow due to fast reactions and long timescales.

\subsubsection{Disadvantages of Continuous Variable Conditions}

The above performance enhancements, along with a careful choice of ODE solver from \textit{DifferentialEquations.jl}, can enable highly efficient ODE solutions over long timescales under challenging environmental conditions (such as high temperatures, which exacerbate the stiffness of generated ODE systems). However, the continuously variable kinetic formalism has two major disadvantages.

First, to compile a continuous analytical RRE that is dependent on environmental conditions, rate constants for every reaction must be expressible purely as a function of those conditions. This limits the level of theory that can be used to describe rate constants, as theories beyond the standard (harmonic/rigid-rotor) transition state theory (TST) approach are too complex to be expressed as continuous functions of external conditions. The continuously-variable kinetics approach is therefore currently limited to an Arrhenius theory-based rate expression.

Second, even when using analytically expressible rate constants with simple functional forms, compiling high-performance functions from symbolic expressions at runtime comes with a high computational cost. Each ODE in the resulting \verb|ODESystem| can contain many terms, each with their own rate constants which must be expanded out to obtain their full functional forms. This leads to an \verb|ODESystem| that can take several hours to compile, which is problematic in the context of the iterative CRN exploration algorithm detailed in Section \ref{sec:networkAssembly_iterative}, where many CRNs must be compiled and solved to reach convergence. This is made even worse by the poor scaling of compilation time with CRN size, whereby every new species adds an ODE to the \verb|ODESystem| and reactions involving such species can appear in many other species' ODEs. All new reactions come with rate constants that must similarly be expanded, causing compilation time to increase dramatically.

\subsection{Simulations with Discrete Variable Conditions}\label{sec:DiscreteVariableKinetics}

For the reasons noted above, it is often impractical to use the \emph{continuous} variable kinetic formalism when solving for the long-timescale kinetics of molecular degradation within large CRNs. We therefore introduce an approximation to this formalism that dramatically cuts down on compilation time with negligible loss in solution accuracy (summarised in Fig. \ref{fig:Modelling_continuous_vs_discrete}b).

Compilation time for RREs with static kinetics is a fraction of that of their continuous variable kinetic counterparts. This is due both to the expansion of every rate constant expression required to form the variable kinetic \verb|ODESystem|, and also to the symbolic analytic partial differentiation that must occur on these complex \verb|ODESystem|s to obtain their analytic Jacobians. As static kinetic RREs only have a single parameter for each rate constant, there is no such expansion and their analytic Jacobians are far simpler to calculate.

These parametric rate constants are usually set at the beginning of a static kinetic simulation and left as such. However, it is possible to modify their values at runtime by using callbacks within the solution of the ODE. \textit{DifferentialEquations.jl} features a rich callback library that facilitates such modifications, allowing us to discretely update rate constants on a finite grid of time points throughout a simulation. 

The method begins with the \verb|ConditionSet|, where users can set a rate constant update timestep $\tau_{r}$. This is used to generate an array of stopping points $\mathbf{t_{\text{stops}}}$ for each variable condition profile along the global simulation timespan, although checks are in place to ensure that no unnecessary stops are made during times when a condition is constant:
\begin{equation}
    \mathbf{t_{stops}^{X}} = \left\{ \tau_{r}n \; \bigg| \; n \in \left\{ 1,2,3,...,\dfrac{t_{\text{end}}}{\tau_{r}}  \right\} ,\textup{abs}\left(\dfrac{dX}{dt}\left( \tau_{r}n \right)\right) > 0 \right\},
    \label{eqn:DiscreteTStops}
\end{equation}
\noindent where $t_{\text{end}}$ is the global simulation end-time and $X$ is an arbitrary variable condition. Once $\mathbf{t_{stops}^{X}}$ has been calculated for each variable condition profile, the arrays are combined into a single set of stopping points for the whole \verb|ConditionSet|, $\mathbf{t_{stops}^{global}}$. 

When the discrete kinetic solver is called, the variable condition profiles in the \verb|ConditionSet| are each independently solved over the simulation timespan, decoupled from the main RRE. This reduces the stiffness of the main RRE, and allows us to pre-calculate the value of each variable condition at the times within $\mathbf{t_{stops}^{global}}$ using \textit{DifferentialEquations.jl}'s solution interpolation. These values are used within the kinetic calculator to calculate the reaction rates at each stopping point, forming a matrix of precalculated rates $\mathbf{K_{precalc}}$. We provide example code showing how $\mathbf{K_{precalc}}$ is populated in Supplementary Figure S2.

During RRE solution, every time point in $\mathbf{t_{stops}^{global}}$ is used as a time-stepping milestone. Once the solver reaches a time point $i$ within $\mathbf{t_{stops}^{global}}$, a callback modifies its internal state by replacing its parameters $\mathbf{k}$ with the precalculated rate constants for that time stop, $\mathbf{K_{precalc}^{i}}$. The simulation then continues with these rate constants until the next time within $\mathbf{t_{stops}^{global}}$ is reached, at which point the callback is repeated.

By discretely updating rate constants on a fine enough $\tau_{r}$ grid, it is possible to recover an approximate concentration-time profile for every species in a given CRN with negligible error when compared to the result generated by the continuous formalism in Section \ref{sec:ContinuousVariableKinetics}. We show that this approximation holds while the gradients of the condition profiles are locally constant in Section S2 of the Supplementary Information. Furthermore, since the compiled \verb|ODESystem| takes a form closer to the static kinetic RRE in Equation \ref{eqn:ToyNetworkStatic} rather than the continuous formalism RRE in Equation \ref{eqn:ToyNetworkContinuousA}, CRN compilation time is dramatically reduced. Full results are discussed in Section \ref{sec:CaseStudy}.

While compilation time is greatly improved under the discrete formalism, we still must also maintain the previously discussed solver optimisations to keep solution times relatively low. This is simple for all optimisations discussed, apart from the multi-timescale approach to simulation time accumulation in Section \ref{sec:ContinuousKinetics_FastReactions}. For this to co-exist with the discrete formalism, care must be taken to convert the global-time $\mathbf{t_{stops}^{global}}$ to local chunk-time values. This is done using a variation of Equation \ref{eqn:tGlobal}:
\begin{equation}
    \mathbf{t_{stops}^{local}} = \left\{ t_{\text{stop}}^{\text{global}} - \tau_{c}n_{c} \; \bigg| \; t_{\text{stop}}^{\text{global}} \in \mathbf{t_{stops}^{global}}, \tau_{c}\left( n_{c}+1 \right) \leq t_{\text{stop}}^{\text{global}} < \tau_{c}\left( n_{c}+2 \right) \right\}.
    \label{eqn:tLocal}
\end{equation}
This selects the global time stops that lie within the current simulation chunk's timespan and converts them into time stops which can be used on the local timespan.

The decoupling of variable rate constant calculation from the main RRE is a vital advantage of the discrete formalism employed here, as it allows for almost any level of rate constant calculation to take place without interrupting the overall CRN solution (provided the rate constants are not themselves dependent on mid-solution quantities such as species concentration). For example, it would be possible to pre-calculate DFT-level TST rate constants based on the experimental conditions requested and to use them on-the-fly within discrete kinetic simulations that mirror the prohibitively expensive exact continuous kinetic simulations at this level.

\subsubsection{Disadvantages of Discrete Variable Conditions}

The discrete kinetic formalism has many benefits, but should not always be used under every circumstance. Changing the rate constants at discrete intervals introduces discontinuities in the concentration gradient of every species simultaneously. These can typically be readily handled by ODE solvers that are capable of adaptive time-stepping, but such solvers must advance simulation time very carefully to maintain numerical stability. This can sometimes slow the overall solution of CRNs compared to the continuous formalism, depending on the severity of the discontinuities. The solution cost of discrete formalism simulations increases as $\tau_r$ is decreased, as this necessitates more stopping points where the callback is run, and therefore more gradient discontinuities. Conversely, $\tau_r$ can be increased to reduce the number of stopping points, although this can make the concentration gradient discontinuities at these points larger and decrease the numerical stability of the solution.

The overall computational time-savings from discrete formalism compilation usually vastly outweigh the potential small performance degradation to solution time. However, one may envisage two situations in which this advantage is not realized:
\begin{enumerate}[label=\emph{\alph*})]
    \item an exact, smooth solution to a given RRE is required,
    \item the gradient discontinuities introduced by the discrete formalism are so severe that solution time is being impacted and the RRE is being repeatedly rerun, for example as part of an ensemble calculation. In this case, the additional cost of continuous formalism compilation may be outweighed by the savings from continuous formalism solving as the \verb|ODESystem| only needs to be compiled once and can be re-solved many times.
\end{enumerate}
\noindent Outside of these situations though, the discrete variable kinetics formalism can be used to greatly accelerate kinetic modelling of generated CRNs; this, as well as the advantages of iterative CRN construction, is demonstrated in the case study below.

\section{Case Study}\label{sec:CaseStudy}

To showcase the functionality implemented in the \textit{Kinetica} packages, we study the CRN and kinetics of ethane pyrolysis under a variable temperature profile. There are a wealth of experimental and theoretical results pertaining to this problem under a variety of variable experimental conditions that make this an excellent system for benchmarking the performance and capabilities of our methods\cite{Susnow1997, Ranzi1997, Tranter2002}. In particular, we aim to replicate some of the experimental pyrolysis results of Xu \latin{et. al.}\cite{Xu2011}, who performed a series of pyrolysis experiments in a tubular quartz flow reactor at temperatures ranging from 550--850$^{\circ}$C, before analysing the pyrolysis products with a combination of gas chromatography (GC), thermal conductivity detection (TCD), flame ionisation detection (FID) and mass spectroscopy (MS). These results are particularly interesting because they include measured temperature profiles along the center line of the flow reactor, which we can replicate with our methodology.

In all of the following results, we employ the \textit{KineticaKPM.jl} modular kinetic calculator, which uses machine learning to predict reaction activation energies on-the-fly with TST. A detailed comparison of the results of our kinetic simulations with experimental pyrolysis results will be presented in an upcoming paper; here, we focus on comparing and contrasting the aspects of CRN construction (direct and iterative methods) and kinetic simulations (continuous and discrete methods) described above.

\paragraph{Variable Temperature Conditions}

To replicate the experimental pyrolysis results in ref. \citenum{Xu2011}, a \verb|ConditionSet| was constructed with simulation temperature bound to a \\\verb|DoubleRampGradientProfile|. This is a gradient-based condition profile that defines two linear temperature ramps separated by a plateau. The relevant gradient function is
\begin{equation}
    \frac{\mathrm{d} X\left( t \right)}{\mathrm{d}t} = 
    \begin{cases}
        0.0, & \text{if } t < t_{r1, \text{start}} \\
        r_1, & \text{if } t_{r1, \text{start}} \leq t < t_{r1, \text{end}} \\
        0.0, & \text{if } t_{r1, \text{end}} \leq t < t_{r2, \text{start}} \\
        r_2, & \text{if } t_{r2, \text{start}} \leq t < t_{r2, \text{end}} \\
        0.0, & \text{if } t \geq t_{r2, \text{end}} 
    \end{cases}
    \label{eqn:Condition_profile_doubleramp}
\end{equation}
\noindent where $r_1$ and $r_2$ are the rates of change of the two linear ramps, and $t_{r1, \text{start}}$, $t_{r1, \text{end}}$, $t_{r2, \text{start}}$ and $t_{r2, \text{end}}$ are the respective start- and end-times of the first and second ramps, determined by the lengths of the starting, middle and ending plateaus.

\begin{figure}[t]
    \centering
    \includegraphics{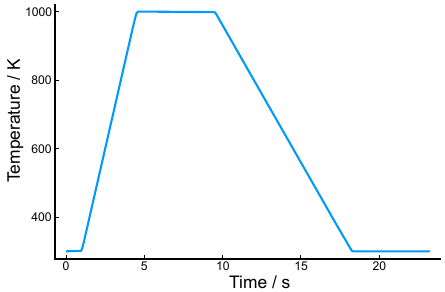}
    \caption{Computational temperature profile matching the 1000 K experimental temperature profile in ref. \citenum{Xu2011}, after implementation and solution within \textit{Kinetica.jl}.}
    \label{fig:CaseStudy_Tprofile}
\end{figure}

By extrapolating the temperature profiles shown along the tubular reactor used in the original work such that the temperature profile starts and ends at 300 K, and by using the reported reactant flow rate to convert distance along the reactor into time, we arrived at the temperature profile shown in Fig. \ref{fig:CaseStudy_Tprofile} to replicate the experimental $725 ^\circ$C ($\approx 1000$ K) pyrolysis profile. The code for the resulting \verb|ConditionSet| is given in Supplementary Figure S3.

\paragraph{SE-GDS Simulations}

The SE-GDS algorithm requires a library of possible graph moves to apply to an input molecular system to stochastically sample chemical reaction space. Here, we used an exhaustive selection of 2-atom and 3-atom moves to explore as much of the available reaction space as possible. The full library of allowed reactions is available in Supplementary Figure S4.

Additionally, SE-GDS requires many input parameters. Except for variables controlling how many reactions are generated per SE-GDS run, these parameters are kept constant for the duration of a CRN generation. They are therefore provided alongside the graph move library as an input file, shown in Supplementary Figure S5.

\subsection{Results of Direct Exploration and Continuous Kinetics}

To perform direct CRN exploration, a system of ethane molecules must be provided to initialize SE-GDS. As high temperature ethane pyrolysis can create free radical species, the radical chain growth mechanism shown in Fig.\ref{fig:Exploration_Ethane_initial_CRN} can occur. The number of molecules in the initial system therefore dictates the maximum size of any species resulting from this mechanism, and thus the complexity of the resulting network. This makes convergence with the direct exploration method increasingly difficult as more molecules are added. At the pyrolysis temperatures being modelled here, species with more than four carbon atoms were not experimentally observed. We therefore chose an initial system of 2 ethane molecules to perform CRN exploration using the direct method.

In \textit{Kinetica}, this calculation is requested by building a \verb|DirectExplore| exploration parameter block, shown in Figure \ref{code:DirectExplore}. This specifies that two ethane molecules (represented as \textit{CC} within SMILES notation) should be repeatedly subjected to single-ended mechanism searches of $n_r = 100$ (defined by the \verb|CDE.radius| parameter) until either $10^{3}$ exploration iterations have elapsed (\verb|maxiters|) or no new reactions have been found for 5 iterations (\verb|rxn_convergence_threshold|). This is a reasonable convergence criterion, requiring $3 \times 10^{3}$  reactions to have been sampled without any change to the CRN to signal convergence (noting that each iteration performs 6 concurrent CDE explorations, as set by \verb|parallel_runs|).

\begin{figure*}[t]
    \centering
    \begin{subfigure}[t]{0.45\textwidth}
        \centering
        \begin{minted}[
            baselinestretch=1.1,
            breaklines,
            linenos,
            numbersep=3pt,
            frame=lines,
            fontsize=\footnotesize,
            framesep=2mm
        ]
        {julia}
exploremethod = DirectExplore(
    rdir_head = "ethane_pyro_1000K_direct",
    reac_smiles = ["CC", "CC"],
    maxiters = 1000,
    rxn_convergence_threshold = 5,
    cde = CDE(
        template_dir = "./cde_template_direct",
        radius = 100,
        parallel_runs = 6
))
        \end{minted}
        \caption{Parameters for direct exploration of an ethane pyrolysis CRN.}
        \label{code:DirectExplore}
    \end{subfigure}\hfill%
    \begin{subfigure}[t]{0.45\textwidth}
        \centering
        \begin{minted}[
            baselinestretch=1.1,
            breaklines,
            linenos,
            numbersep=3pt,
            frame=lines,
            fontsize=\footnotesize,
            framesep=2mm
        ]
        {julia}
pars = ODESimulationParams(
    tspan = (0.0, get_t_final(conditions)),
    u0 = Dict("CC" => 1.0),
    solver = CVODE_BDF(; 
        linear_solver=:KLU
    ),
    solve_chunks = true,
    solve_chunkstep = 0.001,
    progress = true,
    low_k_cutoff = :auto
)
        \end{minted}
        \caption{Parameters for RRE kinetic simulation with chunkwise timestepping.}
        \label{code:SolveParams}
    \end{subfigure}
    \caption{\textit{Kinetica} parameter blocks used within network $C_4^D$ exploration and simulation.}
    \label{code:DirectParams}
\end{figure*}

\begin{figure}
    \centering
    \includegraphics{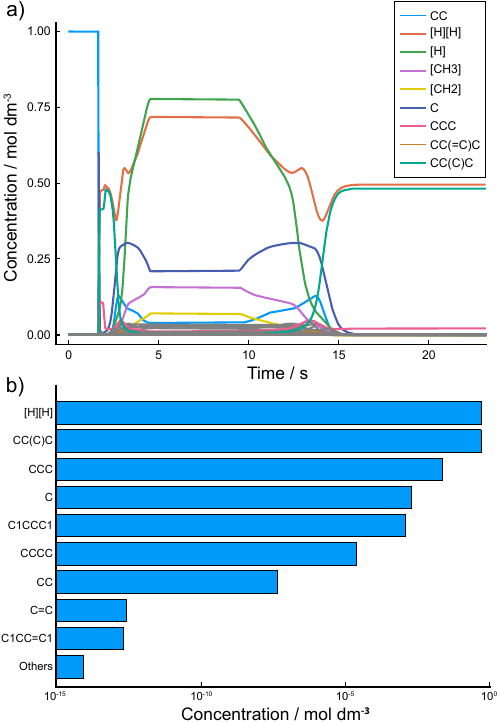}
    \caption{Kinetic simulation results of network $C_4^D$ performed with the continuous rate update approach. a) Concentration-time profiles of prevalent species. b) Final concentrations of species at simulation end.}
    \label{fig:CaseStudy_Continuous_Results}
\end{figure}

This direct exploration resulted in a CRN, which we refer to as network $C_{4}^{D}$, consisting of 3758 reactions and 164 unique species. We performed a kinetic simulation on this CRN using the continuous rate update formalism with chunkwise timestepping at $\tau_c = 1$ ms. This is requested from \textit{Kinetica} by building an \verb|ODESimulationParams| parameter block, shown in Figure \ref{code:SolveParams}. The resulting concentration/time profiles are shown in Fig. \ref{fig:CaseStudy_Continuous_Results}a.

As the simulation temperature increases past 450 K, all ethane (\textit{CC} in SMILES) in the system quickly breaks down, forming methane (\textit{C}) and larger molecules such as propane (\textit{CCC}) and isobutane (\textit{CC(C)C}). The absence of any radical species present in the reaction mixture at this stage indicates preferential recombination into larger species. This conversion to larger species results in fewer hydrogenated carbon atoms on average, noting that diatomic hydrogen (\textit{[H][H]}) is also released. As temperature increases further, these larger species also become unstable and break down into methane, methyl radicals (\textit{[CH3]}) and diradicals (\textit{[CH2]}), and ethane. Hydrogen atoms (\textit{[H]}) are also released in large quantities at higher temperatures. Once the system temperature reaches 1000 K and reaches a plateau, a steady state is quickly formed within the CRN, analogous to a kinetic simulation with static conditions.

Once the temperature starts to decrease again, both single-atom and molecular hydrogen recombine with other species in the system to form stable hydrocarbons such as methane, ethane and propane. Past 725 K however, this behaviour changes towards creating isobutane as the kinetic rate equations being used identify it as the most kinetically-accessible species at ambient temperatures. This leaves nearly all carbon atoms in isobutane molecules at the end of the simulation, as well as the molecular hydrogen that is released as a result. Final species concentrations are shown in Fig. \ref{fig:CaseStudy_Continuous_Results}b.

To demonstrate the poor scaling of the direct exploration method with initial system size, we repeated this exploration with an additional ethane molecule in the starting system, allowing for species containing up to six carbon atoms to be formed. While we do not expect these new species to be formed under the simulation conditions, the direct exploration method does not make any attempt to selectively explore chemical reaction space, instead attempting to find all possible reactions within a given reactive radius $n_{r}$.

This change, while seemingly small, had a profound impact on the resulting CRN (labeled network $C_6^D$). Exploration into the newly-accessible areas of chemical space allowed more than 5000 species to become accessible for reactions, over an order of magnitude larger than the number of species discovered in the CRN comprising just two ethane molecules. The combinatorial nature of CRNs revealed an even larger reaction space, with over $1.34 \times 10^{5}$ reactions being discovered before exploration was stopped by an error. This error was found to be the result of a species being impossible to rationalise into SMILES, another problem associated with uncontrolled exploration into kinetically-inaccessible and uncharted chemical space.

Even if convergence of network $C_6^D$ were possible, CRNs of this size are too expensive to compile into solvable \verb|ODESystem|s under the continuous kinetic formalism described in Section \ref{sec:ContinuousVariableKinetics}; in addition, such CRNs are also extremely inefficient to solve as many reactive pathways are kinetically irrelevant.

\subsection{Validating the Discrete Kinetic Approximation}

While the solution for the kinetic simulation of network $C_4^D$ above was obtainable, the compilation time (and associated memory cost) of such a large network was very high. This and the much larger network $C_6^D$ are therefore ideal benchmarks for the discrete rate-update formalism, as it dramatically reduces this compilation time and makes otherwise inaccessible results obtainable.

\begin{figure*}[t]
    \centering
    \includegraphics{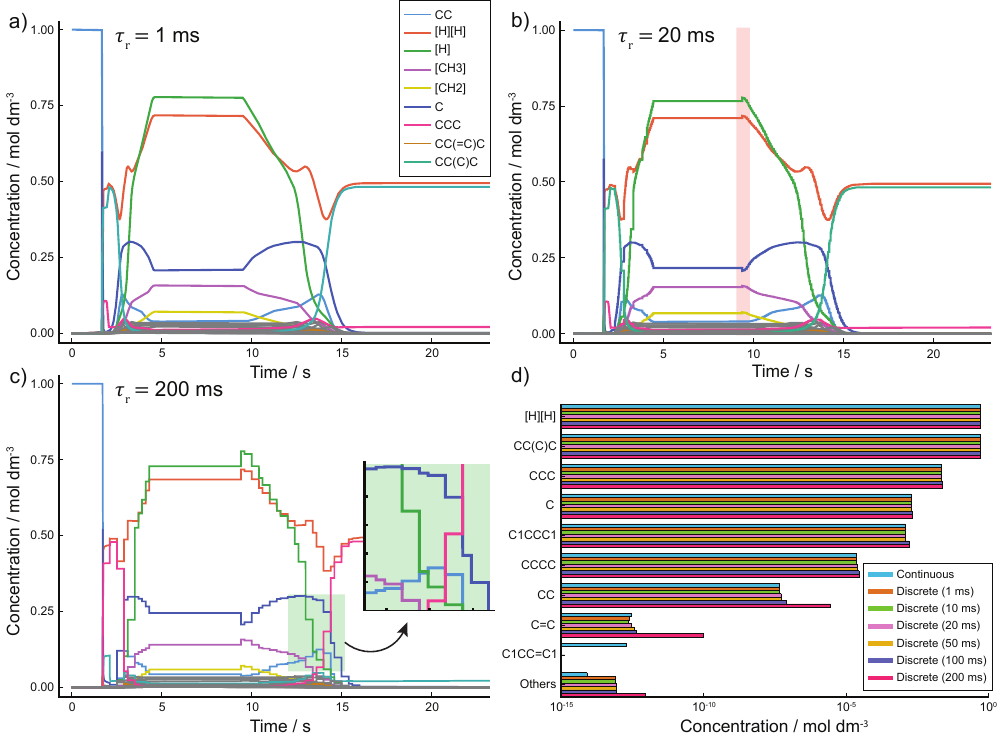}
    \caption{Kinetic simulation results of network $C_4^D$ performed with the discrete rate update formalism. a) Concentration-time profiles with $\tau_r = 1$ ms. b) Concentration-time profiles with $\tau_r = 20$ ms, with concentration anomalies when simulation temperature decreases highlighted. c) Concentration-time profiles with $\tau_r = 200$ ms. d) Comparison between final concentrations of species at the end of continuous and discrete rate update simulations with varying values of $\tau_r$.}
    \label{fig:CaseStudy_Discrete_Results}
\end{figure*}

We first establish that the discrete formalism is a good approximation of fully-continuous variable kinetics by re-running the kinetic simulation of network $C_4^D$ with discretely-updated rate constants. Initially, we update rate constants every $\tau_{r} = 20$ ms. The results of this simulation are shown in Fig. \ref{fig:CaseStudy_Discrete_Results}b.

The $\tau_r = 20$ ms discrete approximation results show a high degree of accuracy with respect to the previous continuous formalism results, both in terms of the final species concentrations at the end of the simulation and the concentration profiles of individual species observed throughout the simulation. The only exception to this is just before $t=10$ s, where the simulation temperature begins to decrease and species concentrations briefly jump to values not observed in the continuous formalism results (marked by a red box in Fig. \ref{fig:CaseStudy_Discrete_Results}b). This is due to the temperature gradient not being locally constant for a short period of time while a discrete rate constant update is taking place.

A direct comparison between the final species concentrations is shown in Fig. \ref{fig:CaseStudy_Discrete_Results}d, where we show that even on a logarithmic scale, the discrete formalism can produce an extremely close match with the continuous formalism results. The exception to this is for species with very small final concentrations, although these species fall into the domain of numerical noise within the simulation and are not guaranteed to be accurate under either rate update formalism. These results were obtained in a significantly shorter time than the continuous formalism results. The RRE compilation time for the kinetic simulation was reduced by more than an order of magnitude, with the RRE solution time being reduced by 55\%. Full results are shown in Table \ref{tab:DiscreteTimings_tau_r}.

Although $\tau_r = 20$ ms was initially chosen as the discrete rate update timestep, we can potentially reduce the solution time of the discrete formalism RRE further by increasing this value. This increases the time between discrete rate constant updates, potentially decreasing the cost of performing these updates. However, this has to be balanced carefully, as increasing $\tau_r$ also increases the relative size of the gradient discontinuities being introduced into the RRE at each update point. These discontinuities require careful adaptive timestepping to resolve, potentially increasing the solution time of the RRE. The effects of modifying $\tau_r$ on the simulation time are also shown in Table \ref{tab:DiscreteTimings_tau_r}, with select kinetic profiles at other values of $\tau_r$ in Figs. \ref{fig:CaseStudy_Discrete_Results}a and \ref{fig:CaseStudy_Discrete_Results}c. Final species concentrations of these simulations are also shown in Fig. \ref{fig:CaseStudy_Discrete_Results}d.

\begin{table*}[ht]
    \centering
    \begin{tabular}{*{4}{|>{\centering\arraybackslash}m{0.9in}}|}
        \hline
        Rate Update Formalism & $\tau_r$ & Compile Time (s) & Simulation Time (s) \\
        \hline
        Continuous & N/A  & 5169 & 326 \\
        \hline
        Discrete & 1 ms & 445 & 240 \\
        Discrete & 5 ms & 449 & 163 \\
        Discrete & 10 ms & 450 & 151 \\
        Discrete & 20 ms & 451 & 145 \\
        Discrete & 50 ms & 451 & 140 \\
        Discrete & 100 ms & 452 & 134 \\
        Discrete & 200 ms & 452 & 135 \\
        \hline
    \end{tabular}
    \caption{Timings for kinetic simulation of network $C_4^D$ under continuous rate update formalism and under discrete rate update formalism with varying $\tau_r$.}
    \label{tab:DiscreteTimings_tau_r}
\end{table*}

For the CRN studied here, we only observed a decrease in RRE solution time as $\tau_r$ was increased (within the margin of error). However, with other CRNs and variable condition profiles we have seen solution times increase dramatically at higher values of $\tau_r$ due to increased numerical instability surrounding the gradient discontinuities. Regardless, while increasing $\tau_r$ significantly can improve solution time, we do not recommend doing so as it becomes very difficult to follow the resulting species concentration profiles and makes simulations prone to numerical error with diminishing gains in solution time. This is especially evident in the low final concentration regime (Fig. \ref{fig:CaseStudy_Discrete_Results}d), where the results with $\tau_r = 200$ ms diverge considerably from those with smaller values of $\tau_r$.

At large values of $\tau_r$, concentration profiles behave in an almost step-wise manner. By zooming in on these concentration steps as in Fig. \ref{fig:CaseStudy_Discrete_Results}c, we can see that the steps are actually very fast changes in concentration, followed by equilibration to a steady state which lasts until the next rate constant update. This behaviour is intuitive when we recall that, between each rate constant update, the RRE has static kinetics and will therefore reach a steady state due to the principle of detailed balance.

\begin{figure}
    \centering
    \includegraphics{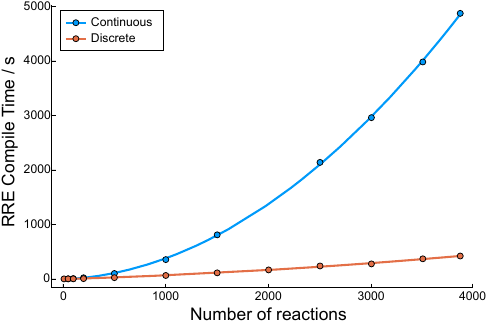}
    \caption{Scaling curves for RRE compilation time under continuous and discrete rate update methods, fit with second order polynomials.}
    \label{fig:DiscreteTimings_scaling}
\end{figure}

We have also investigated the computational scaling of RRE compile time under both rate update formalisms with CRN size by compiling successively larger sub-networks of $C_4^D$. These results are shown in Fig.\ref{fig:DiscreteTimings_scaling}, where a polynomial function was fitted to each set of compile times to enable predictive extrapolation.  The continuous kinetic formalism is shown to scale very poorly with CRN size, while the direct kinetic formalism demonstrates much more favourable scaling behaviour. 

Using these polynomial fits we can approximate how long it would take to compile network $C_6^D$ in its current state. Under the discrete formalism it would take approximately 72 hours which, while a significant amount of time, would not be unobtainable if necessary. However, under the continuous formalism we estimate this RRE would take over 60 days to compile with the current methodology. While this may look fairly dire, we again emphasise that this is one of the expected downfalls of CRN exploration with the direct method, while a more directed approach to exploration of chemical reaction space would remove the need for RREs of this size by reducing the number of reactions considered.

Since the discrete rate update formalism introduces negligible error and greatly expedites the compile time and solve time of RREs (provided a reasonable value of $\tau_r$ is set), all further kinetic simulations in this work will be performed using this strategy. In particular, we use $\tau_r = 10$ ms as this removes the previously-discussed artefacts observed when $\tau_r = 20$ ms and provides excellent accuracy while retaining a fast RRE solution time.

\subsection{Iterative Exploration}

To resolve the systematic under-sampling of generated species in the direct exploration CRN, we repeated the CRN exploration using the iterative exploration method. This method propagates exploration along kinetically-accessible reaction pathways, ensuring that species which exist at high concentrations are reacted and fully sampled under the given simulation conditions.

When performing an iterative exploration, the initial system needs to only contain one of each of the possible initial reactants, as other systems of high concentration seed species are created by \textit{Kinetica} as exploration continues. There is therefore only one ethane molecule requested for reaction in the \verb|IterativeExplore| exploration parameters shown in Fig. \ref{code:IterativeParams}. This parameter block has many of the same options as the simpler \verb|DirectExplore|, except for a few additions. 

The most important of these new parameters is the seed selection concentration cutoff $c_{\text{select}}$ (\verb|seed_conc|), which controls the concentration at which species are selected to be in the seed system at each iterative level of the CRN. Setting this parameter too high results in too few species being selected to undergo further reactions each level, leading to a different kind of under-sampling --- \textit{kinetic} under-sampling. This occurs when the CRN is \textit{systematically} fully sampled (that is, \textit{Kinetica} has explored the CRN until it cannot find any more reactions given the seed species available for reaction), but there are still kinetically important species that remain under-sampled because they have not been selected as seeds. However, setting $c_{\text{select}}$ too low can result in too many seeds being selected for further reactions at each level, making the resulting CRN extremely large. Also of note is the \verb|seed_convergence_threshold| parameter, which dictates how many levels of exploration must be completed with no change in the next level's seed system in order for the network to be considered converged. Usually this can be set to two; after two iterations of exploring reactions exclusively between the same species, the chance of finding any further kinetically important reactions is very low.

\begin{figure}[t]
    \centering
    \begin{minted}[
        baselinestretch=1.1,
        breaklines,
        linenos,
        numbersep=3pt,
        frame=lines,
        fontsize=\footnotesize,
        framesep=2mm
    ]
    {julia}
exploremethod = IterativeExplore(
    rdir_head = "ethane_pyro_1000K_iterative",
    reac_smiles = ["CC"],
    maxiters = 1000,
    rxn_convergence_threshold = 5,
    seed_convergence_threshold = 2,
    seed_conc = 0.02,
    cde = CDE(
        template_dir = "./cde_template",
        radius = 1,
        nrxn = 10,
        parallel_runs = 6
    )
)
    \end{minted}
    \caption{\textit{Kinetica} parameters for iterative exploration used within network $C_4^{I, 0.02}$.}
    \label{code:IterativeParams}
\end{figure}

We began by running an iterative CRN exploration with $c_{\text{select}} = 0.02$ mol dm$^{-3}$. This network was also limited to exploring species containing up to four carbon atoms, but this cannot be achieved by limiting the size of the initial molecular system as in the direct exploration method because the iterative method creates new systems of larger molecules as it proceeds. We therefore use \textit{Kinetica.jl}'s \verb|RxFilter| option to remove reactions resulting in species with greater than four carbon atoms each time a kinetic simulation is run. Further details on how this is implemented are given in Section 3.4 of the Supplementary Information.

We ran the iterative CRN exploration to create network $C_4^{I, 0.02}$. This network required nine levels of iterative exploration to converge, finishing with a network of 3425 reactions and 104 unique species. This is notably smaller than network $C_4^D$ in both reactions and species, and when running a kinetic simulation on the final network the results are markedly different. This is due to network $C_4^{I, 0.02}$ being kinetically under-converged, because kinetically important reactions have not been found due to $c_{\text{select}}$ being set too high.

We therefore performed another iterative CRN exploration with $c_{\text{select}} = 0.01$ mol dm$^{-3}$ to obtain better kinetic convergence. This resulted in network $C_4^{I, 0.01}$, which took 11 levels of exploration to converge, with 8026 reactions and 130 species. This is over double the number of reactions explored in $C_4^D$, but fewer species overall. This occurs because the iterative exploration method never explores species with kinetically inaccessible precursors; as a result, this network is therefore much more thoroughly explored than $C_4^D$.

\begin{figure}[t]
    \centering
    \includegraphics{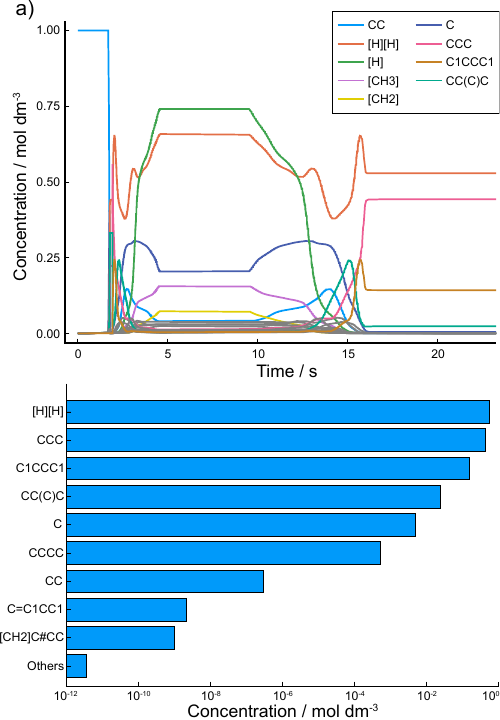}
    \caption{Kinetic simulation results of network $C_4^{I, 0.01}$ performed with the discrete rate update formalism at $\tau_r = 20$ ms. a) Concentration-time profiles of prevalent species. b) Final concentrations of species at the end of simulation.}
    \label{fig:CaseStudy_Cap4_Results}
\end{figure}

The kinetic simulation results of network $C_4^{I, 0.01}$ are shown in Fig. \ref{fig:CaseStudy_Cap4_Results}. While there are many similarities to the kinetic simulation of network $C_4^D$, there are also some differences that clearly demonstrate the presence of other kinetically important reactions. Of note are concentration profiles of propane, isobutane and cyclobutane (\textit{C1CCC1}), the latter of which was found by the direct CRN exploration but never emphasised in the kinetic simulation of $C_4^D$ due to missing reactions.

These additional reactions are mostly observed during the temperature ramps. During heating, a large amount of cyclobutane is temporarily formed, releasing molecular hydrogen, before breaking down as the temperature continues to increase. During cooling, the rise in concentration of isobutane is halted by increasing concentrations of propane and cyclobutane, leading to the latter being the majority products instead of the former. This is accompanied by an increased final concentration of molecular hydrogen, which is released by the formation of cyclobutane before being consumed by formation of the less substituted propane.

Finally, as with the direct exploration method example, we increased the carbon atom limit in the \verb|RxFilter| to six and ran another iterative CRN exploration with $c_{\text{select}}= 0.02$ mol dm$^{-3}$, forming network $C_6^{I, 0.02}$. After 18 levels of exploration, this produced a converged CRN of 14,274 reactions over 707 species. The kinetic simulation results of this network are easily obtainable with the discrete rate update formalism, but they reveal problematic trends.

While not necessarily kinetically converged, the main final product after kinetic simulation of $C_6^{I, 0.02}$ is methyl-cyclopentane, a species containing 6 carbon atoms. We know from the experimental reference results that such species should not be present in the final reaction mixture of an ethane pyrolysis at this temperature, indicating that some of the kinetic rate constants being employed are potentially inaccurate. This is due in part to a missing entropic contribution within the rate constants calculated by \textit{KineticaKPM.jl}, which will be the main focus of an upcoming follow-up paper on the possibility of calculating rate constants on-the-fly using machine learned activation energies.

\section{Conclusions}

In this article, we presented \textit{Kinetica.jl}, a Julia package for automated exploration and kinetic simulation of CRNs under arbitrary variable conditions. In particular, we have discussed many of the features of \textit{Kinetica.jl}, including an iterative method for kinetics-guided exploration of chemical reaction space, a modular kinetic calculator interface allowing symbolic condition profiles to be bound to user-defined rate expressions, and a discrete approximation of continuously variable rate constants that greatly accelerates simulations with negligible loss in accuracy.

By combining these elements into a unified workflow engine powered by the SciML ecosystem\cite{Rackauckas2017} and CDE's new SE-GDS algorithm, defined here, we are able to quickly and efficiently generate large, complex CRNs that are tailored to specific environments, unlocking theoretical insight into complex degradation processes. Through a multi-timescale approach to ODE solution within kinetic simulations, we are able to extend this insight to challenging environmental conditions and long timescales. \textit{Kinetica} has been written with extensibility at its core, allowing users to implement custom ways of calculating rate constants which can vary with user-defined experimental conditions. \textit{KineticaKPM.jl} is one such extension, providing a machine leaning-based kinetic calculator that allows for fast solution of temperature-dependent RREs. The details of this calculator and intricacies of machine-learned rate constants will be the subject of future work.

We have demonstrated \textit{Kinetica}'s CRN generation, kinetic simulation and analysis workflow by assembling a number of CRNs for the 1000 K pyrolysis of ethane, with a variable temperature profile set to match an experimental temperature profile measured along the length of a tubular flow reactor. By performing kinetic simulations on the resulting networks, we have demonstrated the utility of the discrete kinetic formalism we employ and highlighted the importance of selective exploration within combinatorially large chemical reaction spaces. Without iterative guidance by kinetic simulation results, it can be near impossible to obtain CRNs that fully characterise all the possible reaction pathways under a set of variable experimental conditions. With this guidance, however, CRNs can become both more accurate and more efficient to solve, which will be vital if automated long-timescale chemical breakdown simulations are to inform experimental degradation studies.

The iterative exploration method does come with an associated cost. Not only does it have to spend more time thoroughly exploring accessible chemical space, but it also requires repeated kinetic simulations to guide the direction of exploration. The former problem could be addressed by modifying the SE-GDS within CDE to deterministically enumerate all unimolecular and bimolecular reactions between one and two species respectively, rather than stochastically sampling chemical reaction space. This would accelerate CRN exploration further, as there would be no repeated exploration of previously discovered reactions, which currently takes up the majority of exploration time while CDE attempts to find every possible reaction. Meanwhile, kinetic simulations could potentially be accelerated further by inclusion of an explicit multi-timescale integrator that takes advantage of timescale separation within the RRE, but this has not been explored yet within \textit{Kinetica}.

\section*{Code Availability}

\textit{Kinetica.jl} is available on GitHub at \url{https://github.com/Kinetica-jl/Kinetica.jl}. \textit{KineticaKPM.jl} is similarly available on GitHub at \url{https://github.com/Kinetica-jl/KineticaKPM.jl}. Both packages are installable through the Julia language \verb|Pkg| package manager.

\begin{acknowledgement}

The authors acknowledge funding from the Atomic Weapons Establishment, the EPSRC Centre for Doctoral Training in Modelling of Heterogeneous Systems
 (EP/S022848/1) at the University of Warwick, and a UKRI Future Leaders Fellowship (MR/S016023/1 and MR/X023109/1). Computing facilities were provided by the Scientific Computing Research Technology Platform (SCRTP) of the University of Warwick.


\end{acknowledgement}

\begin{suppinfo}

Representative GDS input files and example code-blocks demonstrating aspects of rate calculations and kinetic simulations are given in the Supplementary Information. 

\end{suppinfo}

\bibliography{bibliography}

\newpage
\appendix
\renewcommand{\thesection}{S\arabic{section}}
\renewcommand{\theequation}{S\arabic{equation}}
\renewcommand{\thefigure}{S\arabic{figure}}
\begin{titlepage}
\begin{center}
    \begin{LARGE}
        \textbf{Supplementary Information\\}
    \end{LARGE}
    \vspace{5mm}
    \begin{large}
        \textbf{Predicting long timescale kinetics under variable experimental conditions with Kinetica.jl  \\}
    \end{large}
    \vspace{5mm}
    {\large Joe Gilkes$^{1,2}$, Mark Storr$^3$, Reinhard J. Maurer$^{1,4}$, Scott Habershon$^1$}\\
    \vspace{5mm}
    $^1$Department of Chemistry, University of Warwick, Gibbet Hill Road, CV4 7AL Coventry, UK
    
    $^2$EPSRC HetSys Centre for Doctoral Training, University of Warwick, Gibbet Hill Rd, CV4 7AL, Coventry, UK
    
    $^3$AWE Plc, Aldermaston, UK

    $^4$Department of Physics, University of Warwick, Gibbet Hill Road, CV4 7AL Coventry, UK
    \vspace{5mm}
    UK Ministry of Defence \copyright \, Crown Owned Copyright 2024/AWE
    \vfill
\end{center}
\end{titlepage}

\section{Implementation Examples}

\subsection{Kinetic Calculators}

\begin{figure*}[h!]
    \centering
    \begin{minted}[
        baselinestretch=1.1,
        breaklines,
        linenos,
        numbersep=3pt,
        frame=lines,
        fontsize=\footnotesize,
        framesep=2mm
    ]
    {julia}
# A struct holds all the data that the calculator needs to calculate its rate constants.
mutable struct PrecalculatedArrheniusCalculator{kmType, uType, tType} <: AbstractKineticCalculator
    Ea::Vector{uType}
    A::Vector{uType}
    k_max::kmType
    t_unit::String
    t_mult::tType
end

# Functors of the calculator struct can be called to calculate rate constants with the specified conditions.
function (calc::PrecalculatedArrheniusCalculator{uType, uType, tType})(; T::Number) where {uType, tType}
    k_r = calc.A .* exp.(-calc.Ea / (Constants.R * T)) * Constants.N_A * calc.t_mult
    return 1.0 ./ ((1.0 / calc.k_max) .+ (1.0 ./ k_r))
end

# Lets callers know if a set of symbolic conditions can be used with a calculator.
# This calculator can only handle temperature as a condition, so any other symbolic conditions will make this return false.
function has_conditions(::PrecalculatedArrheniusCalculator, symbols::Vector{Symbol})
    return all([sym in [:T] for sym in symbols])
end
    \end{minted}
    \caption{Shortened definition of the \texttt{PrecalculatedArrheniusCalculator} kinetic calculator in \textit{KineticaCore}.}
    \label{code:Calculator_example}
\end{figure*}

\subsection{Discrete Formalism Rate Precalculation}

\begin{figure*}[h!]
    \centering
    \begin{minted}[
        baselinestretch=1.1,
        breaklines,
        linenos,
        numbersep=3pt,
        frame=lines,
        fontsize=\footnotesize,
        framesep=2mm
    ]
    {julia}
# Accumulate global time stops array from all condition profiles.
tstops = get_tstops(conditions)

sc_symbols, sc_values = get_static_conditions(conditions)
vc_symbols, vc_profiles = get_variable_conditions(conditions)

# Initialise length(tstops) rows of rate constants for all reactions in the CRN.
K_precalc = [zeros(rd.nr) for _ in tstops]

# Iterate over time stops.
for (i, tstop) in enumerate(tstops)
    # Create a mapping between all condition symbols and their respective values/profiles.
    bound_conditions = vcat(
        [symbol => value for (symbol, value) in zip(sc_symbols, sc_values)],
        [symbol => profile(tstop) for (symbol, profile) in zip(vc_symbols, vc_profiles)]
    )

    # Pre-calculate rate constants at each time stop.
    # Elipses separate out bound conditions into keyword arguments to the calculator.
    K_precalc[i] = calculator(; bound_conditions...)
end
    \end{minted}
    \caption{Rate precalculation procedure for discrete variable kinetics simulations. By binding static conditions to their assigned values and variable conditions to the interpolated values at each \texttt{tstop} in $\mathbf{t_{stops}^{global}}$, rate constants can be precalculated and inserted as rows in $\mathbf{K_{precalc}}$.}
    \label{code:Precalculated_rates}
\end{figure*}

\section{Discrete rate update formalism proof}
\label{SI:DiscreteFormalismProof}

For the discrete approximation to hold, the gradient of each rate constant with respect to time in the continuous formalism must be approximately equal to the gradient of each rate constant between two neighbouring rate update points, i.e. for rate update points $t_1$ and $t_2$:

$$
\frac{dk}{dt} \approx \frac{k\left( t_2 \right) - k\left( t_1 \right)}{\Delta t}
$$

\noindent This applies to rate constants given by the equation

$$
k\left( t \right) = \dfrac{1}{\dfrac{1}{k_{max}}+\dfrac{1}{N_{A}\sigma_{AB}\sqrt{\dfrac{8k_{B}T\left( t \right)}{\pi\mu_{AB}}}\exp\left(-\dfrac{E_{a}}{RT\left( t \right)}\right)}}
$$

\noindent There is no continuous analytical expression for $\frac{dT}{dt}$, but it is (usually) locally constant. We will therefore make the approximation that within a local region of the kinetic profile, $\frac{dT}{dt} = \alpha$. For the sake of legibility, we also simplify the rate constant expression by removing all constants except for $E_a$:

$$
k\left( t \right) = \sqrt{T} e^{-\dfrac{E_a}{T}}
$$

\subsection{Continuous Gradient}

\begin{equation*}
\begin{split}
    \frac{dk}{dt} & = \frac{dk}{dT} \cdot \frac{dT}{dt} \\
    \frac{dk}{dT} & = \frac{1}{2T^{\frac{1}{2}}}e^{-\dfrac{E_a}{T}} + \frac{E_a}{T^{\frac{3}{2}}}e^{-\dfrac{E_a}{T}} \\
    & = e^{-\dfrac{E_a}{T}} \left(\frac{1}{2T^{\frac{1}{2}}} + \frac{E_a}{T^{\frac{3}{2}}}\right) \\
    \frac{dT}{dt} & = \alpha \\
    \therefore \frac{dk}{dt} & = \alpha e^{-\dfrac{E_a}{T}} \left(\frac{1}{2T^{\frac{1}{2}}} + \frac{E_a}{T^{\frac{3}{2}}}\right)
\end{split}
\end{equation*}

\subsection{Discrete Gradient}

Assuming $T_1 = T\left( t_1 \right)$ and $T_2 = T\left( t_2 \right)$:

\begin{equation*}
    \begin{split}
        t_2 & = t_1 + \Delta t \\
        \therefore T_2 & = T_1 + \alpha\Delta t \\
    \end{split}
\end{equation*}

\noindent The rate constants at the rate update points therefore take the form:

\begin{equation*}
    \begin{split}
        k\left(t_1\right) & = \sqrt{T_1} \cdot e^{-\dfrac{E}{T_1}} \\
        k\left(t_2\right) &= \sqrt{T_1 + \alpha\Delta t} \cdot e^{-\dfrac{E}{\left( T_1 + \alpha\Delta t \right)}}
    \end{split}
\end{equation*}

\noindent The gradient of the line connecting these two points is therefore:

\begin{equation*}
        \dfrac{k\left( t_2 \right) - k\left( t_1 \right)}{\Delta t} = \dfrac{\sqrt{T_1 + \alpha\Delta t} \cdot e^{-\dfrac{E}{\left( T_1 + \alpha\Delta t \right)}} - \sqrt{T_1} \cdot e^{-\dfrac{E}{T_1}}}{\Delta t}
\end{equation*}

\noindent As $\Delta t$ tends towards zero, i.e. as the rate update timestep gets smaller and the approximation gets better:

\begin{equation*}
    \begin{split}
        \lim_{\Delta t\to 0} \dfrac{k\left( t_2 \right) - k\left( t_1 \right)}{\Delta t} & = \dfrac{\alpha e^{-\dfrac{E_a}{T_1}}\left( 2E_a + T_1 \right)}{2T_{1}^{\frac{3}{2}}}\\
        & = \alpha e^{-\dfrac{E_a}{T_1}} \left( \dfrac{2E_a}{2T_{1}^{\frac{3}{2}}} + \dfrac{T_1}{2T_{1}^{\frac{3}{2}}} \right)\\
        & = \alpha e^{-\dfrac{E_a}{T_1}} \left(\frac{1}{2T_{1}^{\frac{1}{2}}} + \frac{E_a}{T_{1}^{\frac{3}{2}}}\right)
    \end{split}
\end{equation*}

\noindent This is equivalent to the continuous formalism's expression for $\frac{dk}{dt}$, provided we accept that $T_1 = T\left( t \right)$, which is true within the limit imposed. This potentially also only holds while $\frac{dT}{dt}$ is a constant, although this requires further inspection, e.g. when $T_2 = T_1 + \sin\left( t_2 \right) - \sin\left( t_1 \right)$.

\section{Additional information for case study}
\subsection{\texttt{ConditionSet} for pyrolysis temperature profile}

\begin{figure}[h!]
    \centering
    \begin{minted}[
        baselinestretch=1.1,
        breaklines,
        linenos,
        numbersep=3pt,
        frame=lines,
        fontsize=\footnotesize,
        framesep=2mm
    ]
    {julia}
conditions = ConditionSet(Dict(
    :T => DoubleRampGradientProfile(;
        X_start = 300.0,
        t_start_plateau = 1.0,
        rate1 = 200.0,
        X_mid = 1000.0,
        t_mid_plateau = 5.0,
        rate2 = -80.0,
        X_end = 300.0,
        t_end_plateau = 5.0,
        t_blend = 0.1
    )),
    tconvert(20.0, "ms", "s")
))
    \end{minted}
    \caption{Kinetica \texttt{ConditionSet} used within case study of 1000 K ethane pyrolysis. Defines a variable temperature profile that starts with a 1 s plateau (\texttt{t\_start\_plateau}) at 300 K (\texttt{X\_start}), before rising to 1000 K (\texttt{X\_mid}) at a rate of 200 K/s (\texttt{rate1}). Temperature is held at this plateau for 5 s (\texttt{t\_mid\_plateau}) before falling back to 300 K (\texttt{X\_end}) at a rate of 80 K/s (\texttt{rate2}), where temperature is held for 5 s (\texttt{t\_end\_plateau}) until the end of the simulation. A blending time of 0.1 s (\texttt{t\_blend}) is used to smooth out transitions between condition gradient regimes by linear interpolation. A 20 ms discrete rate update timestep is passed to allow the \texttt{ConditionSet} to calculate the time points at which rate constant updates will occur.}
    \label{SI:ConditionSet}
\end{figure}

\subsection{SE-GDS graph move library}

2-atom and 3-atom graph moves for CDE's SE-GDS algorithm used within Kinetica ethane pyrolysis CRN generation. Each graph move is a block starting with \verb|move|, followed by the number of atoms involved in that move $n_m$ (\verb|natom 2| represents a 2-atom move). This is followed by two $n_m \times n_m$ subgraphs; the first represents part of the CM of the current molecule system and the second represents a modification to the first. 

This is followed by a \verb|labels| keyword, which is accompanied by $n_m$ elemental symbols. This can be used to constrict the graph move to a specific set of atoms, where each symbol corresponds to its respective atom in the above subgraphs. Elemental symbols can also take \verb|*| as a value, indicating that any atom which fits the bonding pattern in the subgraph can be used. Finally, an optional \verb|prob| keyword allows for setting a probability of graph move selection. This probability is relative to the other probabilities set within the graph move library.

For example, the first graph move in the below library represents a bond breaking between any two connected atoms. This is defined by a 2-atom move with \verb|labels * *|, so the move is unrestrained by atom types, and only depends on a matching subgraph. This subgraph is defined as two atoms which are connected to each other (a 1 at coordinates $\left( a_1, a_2 \right)$ in the current CM, where $a_1$ and $a_2$ are the indeces of the selected atoms, and also at $\left( a_2, a_1 \right)$ by symmetry). The move specifies that these 1s should be modified to 0s, disconnecting the two selected atoms, with a relative probability of $0.1$ out of a total of $1.6$, i.e. $6.25\%$ of the time.

\begin{Verbatim}[baselinestretch=0.9]
# Curated movelist for ethane.

# Bond breaking: A-B -> A / B
move 
natom 2
-
0 1
1 0
-
0 0
0 0
-
labels * * 
prob 0.1

# Bond making: A / B -> A-B
move
natom 2
-
0 0
0 0
-
0 1
1 0
-
labels * * 
prob 0.3

# H2 dissociation from a single carbon: H-C-H -> H-H / C
move
natom 3
-
0 0 1
0 0 1
1 1 0
-
0 1 0
1 0 0
0 0 0
-
labels H H C 
prob 0.1

# H2 Association to a single carbon: H-H / C -> H-C-H
move
natom 3
-
0 1 0
1 0 0
0 0 0
-
0 0 1
0 0 1
1 1 0
-
labels H H C
prob 0.3

# A-C-B -> A-B / C
move
natom 3
-
0 0 1
0 0 1
1 1 0
-
0 1 0
1 0 0
0 0 0
-
labels * * * 
prob 0.1

# A-B / C -> A-C-B
move
natom 3
-
0 1 0
1 0 0
0 0 0
-
0 0 1
0 0 1
1 1 0
-
labels * * * 
prob 0.3

# A-C / B -> A-B / C
move
natom 3
-
0 0 1
0 0 0
1 0 0
-
0 1 0
1 0 0
0 0 0
-
labels * * * 
prob 0.2

# A-B / C -> A-C / B
move
natom 3
-
0 1 0
1 0 0
0 0 0
-
0 0 1
0 0 0
1 0 0
-
labels * * * 
prob 0.2

# A-B-C -> C-A-B
move
natom 3
-
0 1 0
1 0 1
0 1 0
-
0 1 1
1 0 0
1 0 0
-
labels * * *
prob 0.2
\end{Verbatim}
\begin{figure}[h!]
\caption{SE-GDS graph move library used in case study of 1000 K ethane pyrolysis.}
\label{SI:GraphMoveLibrary}
\end{figure}

\subsection{SE-GDS parameters}

CDE input file for SE-GDS mechanism generation in direct CRN exploration of 1000 K ethane pyrolysis. Parameter definitions are documented in CDE's code repository at \url{https://github.com/HabershonLab/cde}.

Important parameters include:
\begin{itemize}
    \item \verb|startfile| - defines the geometry of the initial molecule system used for this SE-GDS run. This is usually generated by Kinetica when used within a CRN generation workflow.
    \item \verb|valencerange| - defines acceptable valences for atom types. In this input, carbon atoms must be connected to 2-4 other atoms. Hydrogen atoms can be connected to 0 or 1 other atoms, allowing for hydrogen radical formation.
    \item \verb|reactiveatomtypes| - defines atom types that are allowed to react.
    \item \verb|nmcrxn| - number of mechanisms to generate, each of length \verb|nrxn|.
    \item \verb|nrxn| - number of reactions to sample per generated mechanism.
\end{itemize}

\verb|nmcrxn| and \verb|nrxn| are usually controlled by settings within Kinetica by the parameters \verb|nrxn| and \verb|radius| respectively within a CDE interface struct. When using the iterative exploration method, changing the latter two values creates CDE inputs with the former two values modified.

\begin{Verbatim}[baselinestretch=0.9]
calctype breakdown
minmolcharge 0      
maxmolcharge 0 
nchargemol 0
maxstepcharge 0
maxtotalcharge 0
optaftermove .true.
ignoreinvalidgraphopt .false.
doinitialopt .false.
pesfull .false.
startfile Start.xyz

stripinactive .true.
optendsbefore .true.
optendsduring .false.

dofconstraints 0
atomconstraints 0

pestype xtb
pesfile xtb.head
pesopttype xtb 
pesoptfile xtb.head
pesexecutable xtb --iterations 1000 --grad
pesoptexecutable xtb --input xtb.inp --iterations 1000 --etemp 1000 --opt tight --grad

movefile moves_2+3.in
gdsthresh 0.5 
gdsspring 0.05
gdsrestspring 0.05
nbstrength 0.04
nbrange 2.5
kradius 0.05
ngdsrelax 10000
gdsdtrelax 0.1
gdsoutfreq 10
graphfunctype 4

valencerange{
C 2 4
H 0 1
}

reactiveatomtypes{  
C 
H 
}

reactiveatoms{  
all
}

reactivevalence{
}

fixedbonds{
}

allowedbonds{
} 

nmcrxn 1
nrxn 100
\end{Verbatim}
\begin{figure}[h!]
\caption{SE-GDS parameters used in case study of 1000 K ethane pyrolysis.}
\label{SI:CDEInput}
\end{figure}

\subsection{\texttt{RxFilter} Implementation}

The \verb|RxFilter| is a data-type within \verb|KineticaCore| that can be passed into a CRN solution call along with the current CRN and a set of simulation parameters. The \verb|RxFilter| allows for definition of an array of functions, each of which takes the current CRN (\verb|sd| and \verb|rd| representing instances of \verb|SpeciesData| and \verb|RxData| respectively) as its arguments and returns a mask of reactions to keep or remove. When a kinetic simulation is requested, each function's mask is calculated and accumulated into a global mask which removes the filtered reactions from the CRN. The filter used for removing reactions creating species with greater than four carbon atoms is shown in Figure \ref{SI:RxFilter}.

\begin{figure}[h]
    \centering
    \begin{minted}[
        baselinestretch=1.1,
        breaklines,
        linenos,
        numbersep=3pt,
        frame=lines,
        fontsize=\footnotesize,
        framesep=2mm
    ]
    {julia}
function large_filter(sd, rd)
    mask = [false for _ in 1:rd.nr]
    for (i, prod) in enumerate(rd.prods)
        for pspec in prod
            if count(j->(j=='C'), pspec) > 4
                mask[i] = true
                break
            end
        end
    end
    return mask
end
filter = RxFilter([large_filter])
        \end{minted}
        \caption{Creation of an \texttt{RxFilter} for removing reactions which create species with greater than 4 carbon atoms.}
        \label{SI:RxFilter}
\end{figure}

\end{document}